\pdfoutput=1

\documentclass[preprint,fleqn,5p,numbers,sort&compress]{elsarticle}

\usepackage{graphicx}
\usepackage{amssymb,amsmath}
\usepackage{natbib}
\usepackage{hyperref}
\hypersetup{pdftitle = The title of my PDF, pdfauthor = My name, pdfsubject= The subject, pdfkeywords = keyword1 keyword2 keyword3}
\hypersetup{colorlinks = true, linkcolor = blue, anchorcolor = red, citecolor = blue, filecolor = red, pagecolor = red, urlcolor = blue}

\journal{International Journal of Non-Linear Mechanics}

\begin{document}

\begin{frontmatter}

\title{Orbit classification and networks of periodic orbits in the planar circular restricted five-body problem}

\author[eez]{Euaggelos E. Zotos\corref{cor1}}
\ead{evzotos@physics.auth.gr}

\author[kep]{K. E. Papadakis}

\cortext[cor1]{Corresponding author}

\address[eez]{Department of Physics, School of Science,
Aristotle University of Thessaloniki, GR-541 24, Thessaloniki, Greece}

\address[kep]{Department of Civil Engineering, Division of Structural Engineering,
University of Patras, 26504 Patras, Greece}

\begin{abstract}
The aim of this paper is to numerically investigate the orbital dynamics of the circular planar restricted problem of five bodies. By numerically integrating several large sets of initial conditions of orbits we classify them into three main categories: (i) bounded (regular or chaotic) (ii) escaping and (iii) close encounter orbits. In addition, we determine the influence of the mass parameter on the orbital structure of the system, on the degree of fractality, as well as on the families of symmetric and non-symmetric periodic orbits. The networks and the linear stability of both symmetric and non-symmetric periodic orbits are revealed, while the corresponding critical periodic solutions are also identified. The parametric evolution of the horizontal and the vertical linear stability of the periodic orbits is also monitored, as a function of the mass parameter.
\end{abstract}

\end{frontmatter}

\section{Introduction}
\label{intro}

One of the most well-studied problems in non-linear dynamics is the restricted few-body problem. This problem describes the motion of a test particle in the combined gravitational field of $N$ primary bodies, with $N \geq 3$. In principle, the test particle may (i) move in bounded (regular or chaotic) orbits around the primaries, (ii) escape from the system, (iii) or even collide with one of the primaries. Knowing the final state of the test particle is of great interest because it allows us to divide the phase space into different areas of motions, which are usually known as basins of escape or bounded basins.

Over the years, several numerical methods and computational tools have been used for determining the character of orbits in the restricted three-body problem (RTBP) \cite{S67}, as well as in other more complicated dynamical systems. These numerical techniques are mainly based on the computation of dynamical indicators, such as the Fast Lyapunov Indicator (FLI) (see e.g., \cite{FLG97,FLFF02,LG16}), the normally hyperbolic invariant manifolds (NHIMs), associated with the equilibrium points of the system (see e.g., \cite{JM99,KLMR01,GKL04,AEL17,LG16,BGMO16,GL18,JZ16,ZJ17}), or even the second species solutions (see e.g., \cite{FNS02}).

The first classification of the solutions of the RTBP has been presented in \cite{C32}, while for the case with $h < 0$ the Chazy classification is supplemented by a fundamental hypothesis (see e.g., \cite{A99}). In \cite{N04,N05} the orbital dynamics of the planar RTBP has been investigated by classifying into several categories large sets of initial conditions of orbits. This work has been continued in \cite{Z15d}, while in \cite{Z15a} we used his numerical techniques in an attempt to reveal the orbital structure of the circular restricted problem of three bodies, when one of the primary bodies is an oblate spheroid. In the same vein, in \cite{Z16} we numerically explored the orbital dynamics in the planar circular equilateral restricted four body-problem. The same numerical methods can also be applied for the study of realistic dynamical systems, such as planetary and celestial systems. For example, the orbit classification has been deployed in \cite{dAT14} for the Earth-Moon system, in \cite{QXQ17} for the Sun-Earth-Moon system, in \cite{Z15b} for the Saturn-Titan system and in \cite{Z15c} for the Pluto-Charon system.

The periodic solutions of a dynamical system is, without any doubt, an issue of paramount importance. All started with the pioneer works of Poincar\'{e} \cite{P92}, who suggested that the most efficient method for revealing the orbital structure of a dynamical system is by using the ``valuable solutions" (les solutions pr\'{e}ciuses), as he was calling the periodic solutions. Poincar\'{e} argued, and later it was confirmed by himself \cite{P12} as well as by Birkhoff \cite{B13}, that the periodic solutions are ``dense" everywhere in the set of all bounded solutions.

Since then, a large amount of research papers have been devoted on the study of periodic solutions, in numerous of dynamical systems. In the present work, we will numerically investigate the periodic solutions in the case of the planar restricted five-body problem, for several values of the mass parameter $\beta$. A previous study on the periodic solutions in this dynamical system has been presented in \cite{MPPD02}, while in \cite{PK07} we numerically explored the orbital dynamics of the five-body system, when all the primaries are sources of radiation. Over the years, the problem of five bodies has been proved a very fertile research ground (e.g., \cite{ARCL16,GYS17,Ha05,HJ11,K98,KMV11,LM08,LS09,MC13,M96,R99,SKS16,SSK14}). A generalization of the five-body problem has been performed in \cite{K98} where the motion of the massless test particle, under the influence of $N-1$ primary bodies in a circular configuration, has been investigated.

The present study will be focused on the simplest periodic solutions of the system, that is the periodic orbits which intersect only twice, per period, the horizontal $x$-axis of the system. In particular, we are going to reveal the network of the families of both symmetric and non-symmetric periodic orbits, for several values of the mass parameter. For complicity, for all families of the periodic solutions and for each periodic orbit the corresponding linear stability (horizontal and vertical) will be also computed. At the same time, the horizontal and vertical-critical periodic solutions of all the orbital families will complete our numerical exploration on the network of periodic orbits.

The article's layout is the following: all the important properties of the mathematical system are described in Section \ref{dyn}. The following Section \ref{clas} contains all the numerical outcomes of the orbit classification, while in Section \ref{nets} we present a thorough and systematic numerical analysis on the network on both symmetric and non-symmetric periodic solutions. Our paper ends with section \ref{conc}, where the concluding remarks are given.

\section{Properties of the dynamical model}
\label{dyn}

\begin{figure}[!t]
\centering
\resizebox{\hsize}{!}{\includegraphics{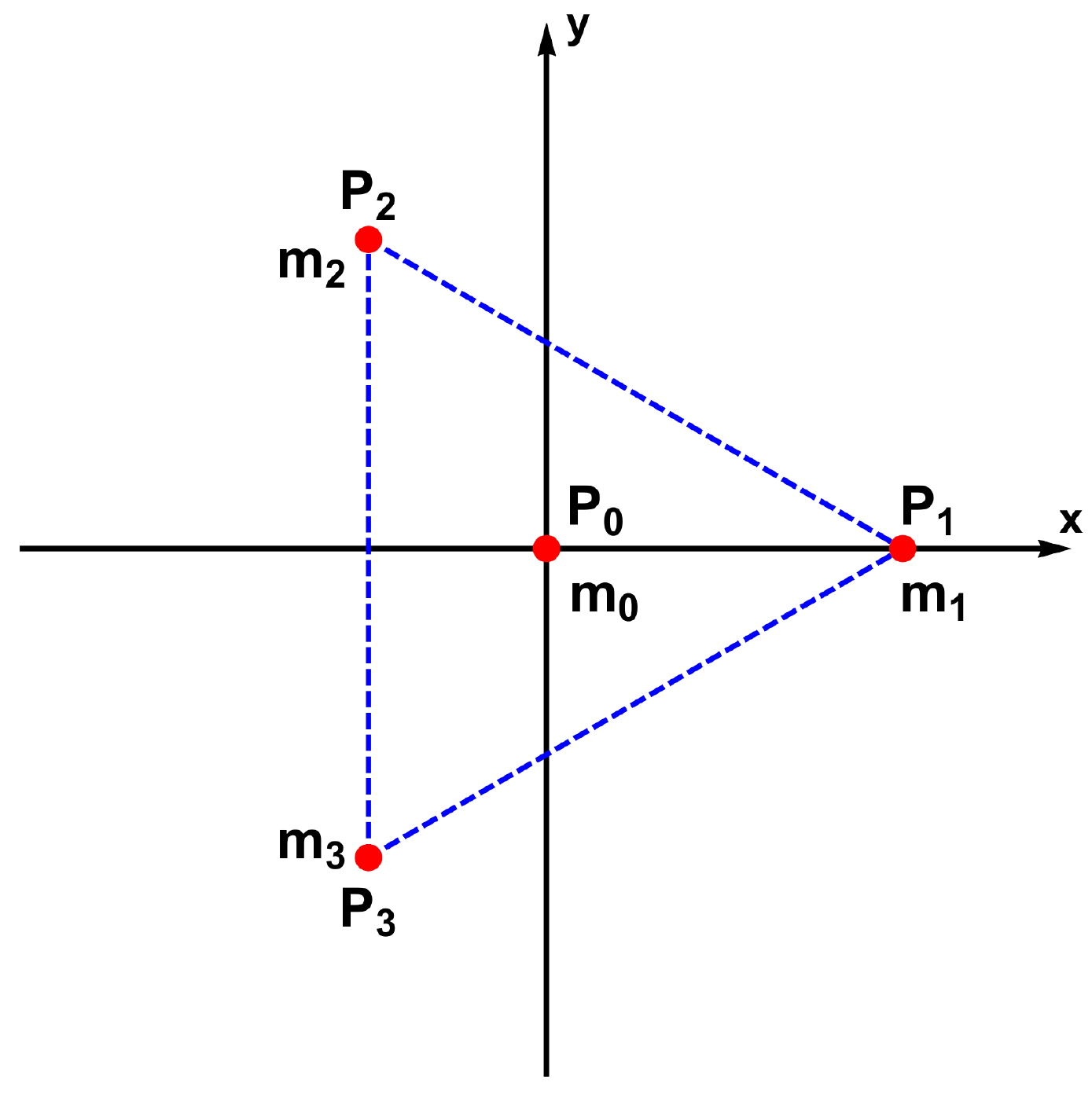}}
\caption{Schematic of the planar configuration of the circular restricted problem of five bodies. Red dots are used for pinpointing the centers of the four primaries, while with blue, dashed lines we indicate the equilateral triangle which is formed by the three equally masses peripheral primaries. (Color figure online).}
\label{conf}
\end{figure}

The configuration of the circular restricted problem of five bodies (four primaries plus a test particle) is the following: three primary bodies, $P_1$, $P_2$, and $P_3$, are situated at the vertices of an equilateral triangle (for which the lengths of the sides are equal to unity), while the fourth $P_0$ lies at the center of the equilateral triangle (see Fig. \ref{conf}). It is assumed, that all the primaries perform circular orbits around their common center of gravity. Furthermore, the fifth body acts as a test particle (with mass significantly smaller that those of the primary bodies) and therefore it does not perturb, in any manner, the coplanar motion of the primaries.

In order to describe the motion of the test particle, we choose a rotating reference frame for which the origin $O(0,0)$ of the coordinates, is identical to the mass center of the system. The first three primaries have equal masses $m_1 = m_2 = m_3 = m$, while the mass of the central primary $P_0$ is $m_0 = \beta~m$. At this point, it should be emphasized that when the central primary body $P_0$ is absent (that is when $\beta = 0$) the configuration of the system is automatically reduced to that of the equilateral restricted problem of four bodies. Due to the equality of the mass of the peripheral primaries ($P_1$, $P_2$, and $P_3$) the system admits a $2\pi/3$ symmetry, with three axes of symmetry $y = 0$, $y = \sqrt{3}$ and $y = -\sqrt{3}$. The centers of the primaries are: $(x_1,y_1,z_1) = (1/\sqrt{3},0,0)$, $(x_2,y_2,z_2) = (-x_1/2,1/2,0)$, $(x_3,y_3,z_3) = (x_2,-y_2,z_2)$, and $(x_0,y_0,z_0) = (0,0,0)$.

The time-independent effective potential is, according to \cite{O88}
\begin{equation}
\Omega(x,y,z) = \frac{1}{k} \sum\limits_{i=0}^3 \left(\frac{m_i}{r_i}\right) + \frac{1}{2}\left(x^2 + y^2 \right),
\label{pot}
\end{equation}
where
\begin{equation}
k = 3\left(1 + \beta~\sqrt{3}\right),
\label{par}
\end{equation}
while
\begin{equation}
r_i = \sqrt{\left(x - x_i\right)^2 + \left(y - y_i\right)^2 + \left(z - z_i\right)^2}, \ \ \ i = 0,...,3,
\label{dist}
\end{equation}
are the respective distances of the test particle from the four primary bodies.

In the synodic frame of reference the set of the equation which describe the motion of the test particle under the mutual gravitation attraction of the four primaries is
\begin{align}
\ddot{x} &- 2 \dot{y} = \frac{\partial \Omega}{\partial x} = \Omega_x(x,y,z), \nonumber\\
\ddot{y} &+ 2 \dot{x} = \frac{\partial \Omega}{\partial y} = \Omega_y(x,y,z), \nonumber\\
\ddot{z} &= \frac{\partial \Omega}{\partial z} = \Omega_z(x,y,z),
\label{eqmot}
\end{align}
where
\begin{align}
\Omega_x(x,y,z) &= - \frac{1}{k} \sum\limits_{i=0}^3 \left(\frac{m_i\left(x - x_i\right)}{r_i^3}\right) + x, \nonumber\\
\Omega_y(x,y,z) &= - \frac{1}{k} \sum\limits_{i=0}^3 \left(\frac{m_i\left(y - y_i\right)}{r_i^3}\right) + y, \nonumber\\
\Omega_z(x,y,z) &= - \frac{1}{k} \sum\limits_{i=0}^3 \left(\frac{m_i\left(z - z_i\right)}{r_i^3}\right),
\label{der1}
\end{align}
are the first order derivatives of the effective potential.

The dynamical system admits only the well-known Jacobi integral of motion. The corresponding Hamiltonian is given by
\begin{equation}
J(x,y,,z\dot{x},\dot{y},\dot{z}) = 2\Omega(x,y,z) - \left(\dot{x}^2 + \dot{y}^2 + \dot{z}^2 \right) = C,
\label{ham}
\end{equation}
where of course $\dot{x}$, $\dot{y}$, and $\dot{z}$ are the respective velocities, associated with the coordinates $x$, $y$, and $z$, respectively, while $C$ is the conserved numerical value of the Hamiltonian.

The number of the equilibrium points in the planar circular restricted five-body problem is in fact a function of the mass parameter $\beta$. In \cite{ZS18} we showed that when $\beta$ lies in the interval $(0,0.01402112)$ there exist fifteen libration points, while when $\beta \in (0.01402113, \infty)$ there are only nine libration points. In the same paper we also investigated the parametric variation of the positions as well as of the linear stability of the equilibrium points, as a function of the mass parameter.

\section{Orbit classification}
\label{clas}

The dynamical system of the planar circular restricted five-body problem has two free parameters, that is the mass parameter $\beta$ and the Jacobi constant $C$. In this section, we will demonstrate how the variation of the numerical values of these parameters influences the orbital dynamics of the system.

Following the pioneer works of Nagler \cite{N04,N05} we will numerically integrate large sets of initial conditions in order to determine the nature of the orbits. The most convenient two dimensional plane is the $(x,C)$ plane which allow us to examine a continuous spectrum of energy levels (note that the orbital energy is directly related with the Jacobi constant through the relation $C = -2E$). Therefore, we define dense uniform grids of $1024 \times 1024$ initial conditions $(x_0,C_0)$, with $-6 \leq x_0 \leq 2.5$ and $-6 \leq C_0 \leq 6$. We choose the particular interval $-6 \leq C_0 \leq 6$ because our previous experience suggests that the most interesting orbital dynamics of the system appears in this energy interval. For all orbits $y_0 = \dot{x_0} = 0$, while the initial value of the $\dot{y}$ velocity is always obtained through the Jacobi constant (\ref{ham}) as $\dot{y_0} = \sqrt{2\Omega(x_0,0) - C_0} > 0$.

The initial conditions of the orbits will be classified into three main categories:
\begin{enumerate}
  \item Orbits performing bounded orbits around the primaries.
  \item Orbits leading to close encounter with one of the primary bodies.
  \item Orbits escaping from the scattering region of the system.
\end{enumerate}

For distinguishing between the above-mentioned types of the orbits we need to define appropriate numerical criteria. In particular, bounded and escaping motion can be obtained as follows: a disk with radius $R = \sqrt{x^2 + y^2} = 10$ is defined on the configuration $(x,y)$ plane. Then if an orbit stays confined, during the entire period of the numerical integration, inside this disk we have the case of bounded motion. On the other hand, if the test particle intersects the radius of that disk, with velocity pointing outward, then we have the scenario of escaping motion. For the case of close encounter motion we define a disk around each primary with close encounter radius $R_{\rm ce} = 10^{-3}$. Therefore close encounter occurs when the test particle crosses the radius $R_{\rm ce}$, with velocity pointing inwards (with direction pointing to the center of the primary). The numerical values regarding both the escaping as well as the close encounter radii were chosen according to \cite{N04,N05}. In the case where a close encounter orbit enters a region of radius $10^{-2}$ around one of the primary bodies we automatically apply the Lemaitre's global regularization method, thus following the approach used in \cite{dAT14}.

It would be very informative if we could further distinguish between ordered and chaotic bounded motion, using a chaos indicator. Fortunately, there are plenty available chaos indicators, such as the Lyapunov Characteristic Exponent (LCE) \cite{BGGS80}, the Fast Lyapunov Indicator (FLI) \cite{FGL97}, the Relative Lyapunov Indicator (RLI) \cite{SES04}, the Orthogonal Fast Lyapunov Indicator (OFLI) \cite{FLFF02}, the Smaller Alignment Index (SALI) \cite{S01}, and the Mean Exponential Growth of Nearby Orbits (MEGNO) \cite{CGS03}. For this task we choose the SALI method. The character of an orbit is revealed through the value of SALI at the end of the numerical integration. More precisely, an orbit is non-escaping regular if SALI $> 10^{-4}$, while an orbit is surely trapped chaotic if SALI $< 10^{-8}$. If however, the final value of SALI lies in the interval $[10^{-8}, 10^{-4}]$ then the particular orbit is called ``sticky orbit" and further numerical integration (using higher integration time) is needed for reaching to safe conclusions about its nature.

In our computations, all the initial conditions of the orbits were numerically integrated for $10^{4}$ time units, using a variable time step. Such a high total time of the numerical integration is justified if we take into account that we want to make sure that all initial conditions will be integrated for sufficient time, so as to reveal their true nature.

\begin{figure*}[!t]
\centering
\resizebox{\hsize}{!}{\includegraphics{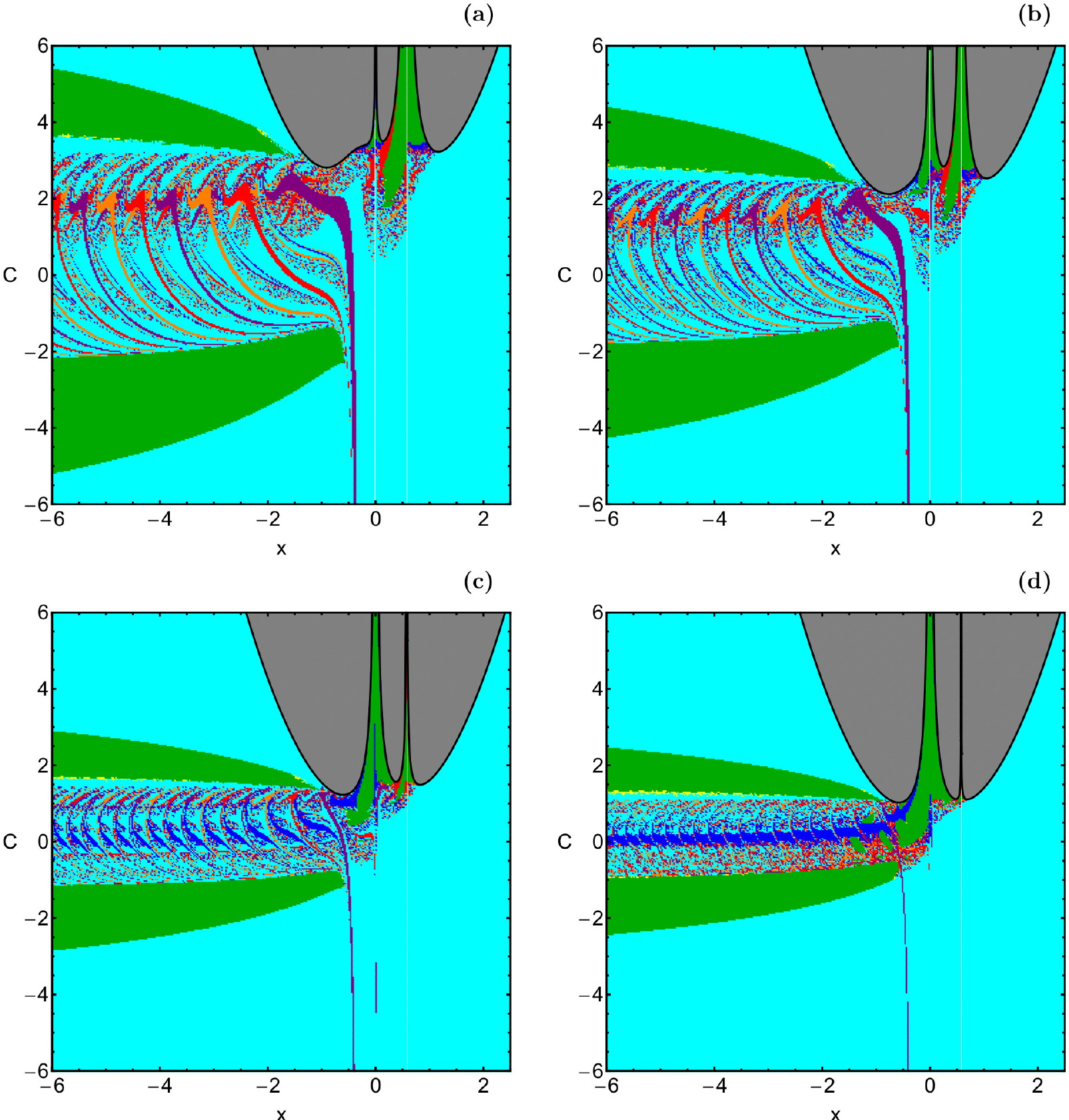}}
\caption{Color-coded diagrams depicting the orbital structure of the $(x,C)$ plane when (a-upper left): $\beta = 0.05$; (b-upper right): $\beta = 0.5$; (c-lower left): $\beta = 5$; (d-lower right): $\beta = 50$. The color code is the following: bounded regular orbits (green); trapped sticky orbits (magenta); trapped chaotic orbits (yellow); close encounter orbits to primary $P_0$ (blue); close encounter orbits to primary $P_1$ (red); close encounter orbits to primary $P_2$ (purple); close encounter orbits to primary $P_3$ (orange); escaping orbits (cyan). The energetically forbidden regions are shown in gray. (Color figure online).}
\label{xC}
\end{figure*}

\begin{figure*}[!t]
\centering
\resizebox{\hsize}{!}{\includegraphics{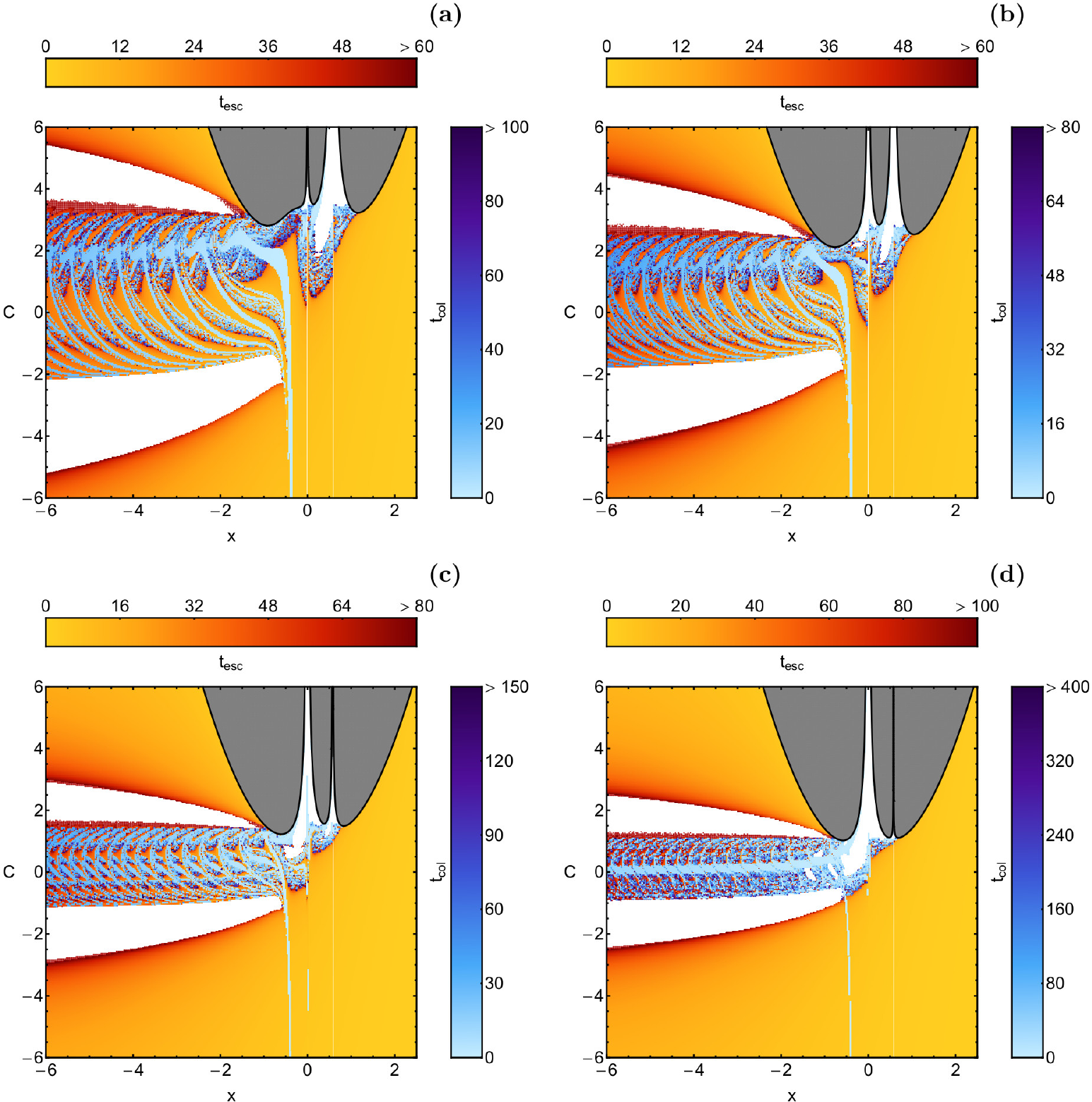}}
\caption{Distributions of the corresponding close encounter and escape times of the orbits, on the $(x,C)$ plane, for the values of the mass parameter $\beta$ of Fig. \ref{xC}(a-d). White color is used for all types of bounded initial conditions (chaotic, sticky and regular). (Color figure online).}
\label{xCt}
\end{figure*}

\begin{figure*}[!t]
\centering
\resizebox{\hsize}{!}{\includegraphics{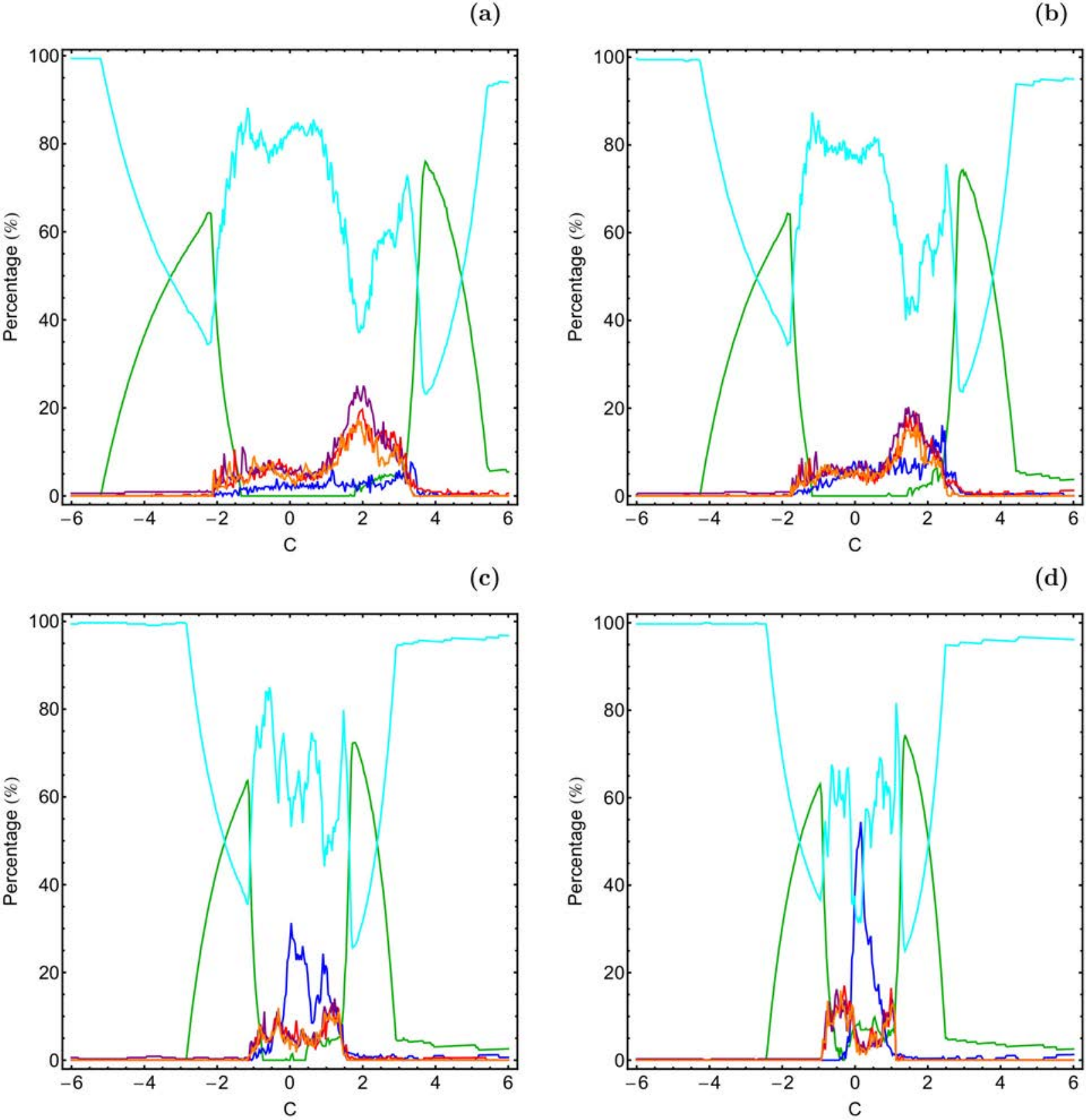}}
\caption{Parametric evolution of the percentages of all types of the orbits with initial conditions on the $(x,C)$ plane, as a function of $C$. (a-upper left): $\beta = 0.05$; (b-upper right): $\beta = 0.5$; (c-lower left): $\beta = 5$; (d-lower right): $\beta = 50$. The color code of the lines is the same as in Fig. \ref{xC}. (Color figure online).}
\label{pxC}
\end{figure*}

The nature of motion on the $(x,C)$ plane, for four values of the mass parameter $\beta$ is presented in Fig. \ref{xC}(a-d), by using modern color-coded diagrams. Each pixel in these plots corresponds to an initial condition $(x_0,C_0)$ and it is colored according to the character of the orbit. The limiting curves, which distinguish between the energetically allowed and forbidden regions of motion are defined as
\begin{equation}
2\Omega(x,y = 0) = C.
\label{zvc}
\end{equation}
and they are shown as black, solid lines in Fig. \ref{xC}(a-d).

It is seen, that between the energetically forbidden regions, and close to the primaries, several stability islands are present. The initial conditions inside these stability islands correspond to both prograde and retrograde quasi-periodic orbits. It should be noted that all these orbits circulate only around one of the primary bodies. Non-escaping regular motion is also observed outside the energetically forbidden areas, where we identify two main stability islands. Additional numerical calculations suggest that the non-escaping regular orbits of these stability islands have a major difference with respect to those located in the interior region (between the forbidden areas). More precisely, all the non-escaping regular orbits with initial conditions in the exterior region circulate around all the primary bodies. Initial conditions corresponding to trapped (sticky or chaotic) motion are located in the vicinity of the stability islands.

\begin{figure*}[!t]
\centering
\resizebox{\hsize}{!}{\includegraphics{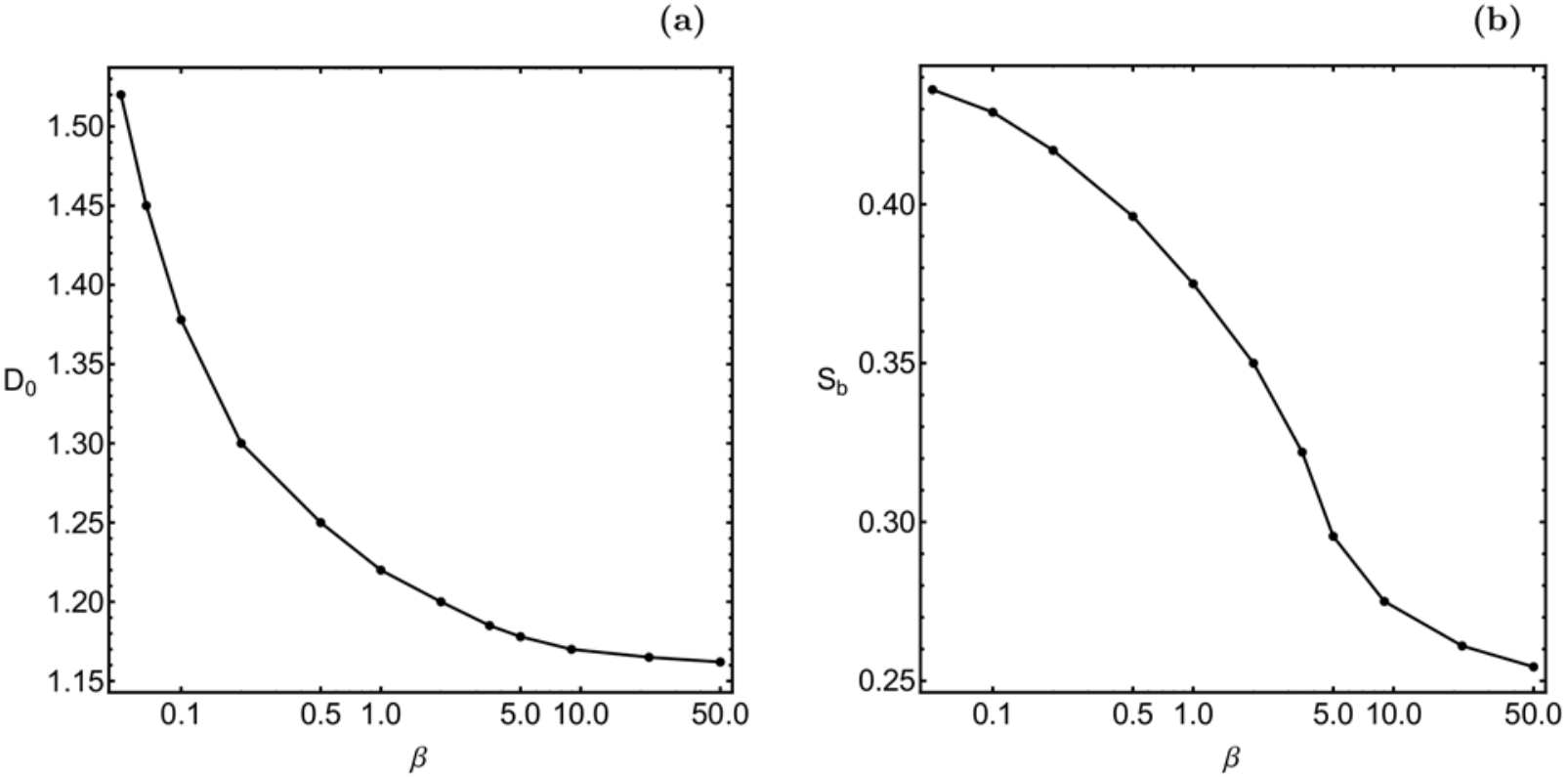}}
\caption{Parametric evolution of the (a-left): uncertainty dimension $D_0$ and (b-right): basin entropy $S_b$ of the $(x,C)$ plane, as a function of the mass parameter $\beta$. It should be noted that a logarithmic scale is used for the horizontal axis.}
\label{frac}
\end{figure*}

The vast majority of the $(x,C)$ planes is covered by initial conditions which escape from the scattering region. Collision initial conditions are mainly located between the two stability islands of the exterior region, where they form complicated spiral structures. One may observe, that between the stability islands of the exterior region there a complicated highly fractal (chaotic) mixture of initial conditions of escaping and close encounter orbits. It should be emphasized that when we state that the particular region on the $(x,C)$ plane is highly fractal, we simply imply that it displays a fractal-like geometry. Later however, we are going to present additional quantitative results regarding the degree of fractality of the $(x,C)$ plane.

With increasing value of the mass parameter $\beta$ the most noticeable changes on the structure of the $(x,C)$ plane are the following:
\begin{enumerate}
  \item The area of the energetically forbidden regions increases, while at the same time these regions tend to converge with each other.
  \item The two stability islands of the exterior region come close to each other and consequently the contained fractal mixture of initial conditions is significantly reduced.
  \item The percentages of the close encounter orbits corresponding to the peripheral bodies ($P_1$, $P_2$, $P_3$) decreases, while the rate of the close encounter orbits to the central primary $P_0$ increases.
\end{enumerate}

The respective distributions of the escape (using tones of red) and close encounter (using tones of blue) times of the orbits are illustrated in Fig. \ref{xCt}(a-d). It becomes evident that the highest values of both escape and close encounter time are encountered in the vicinity of the basin boundaries, where all the fractal structures are present.

Additional useful conclusion can be extracted by monitoring the evolution of the rates of all types of the orbits as a function $C$. In Fig. \ref{pxC}(a-d) we present the parametric variation of the rates of all types of the orbits for the corresponding cases (values of $\beta$), shown earlier in Fig. \ref{xC}(a-d). We see, that in general terms the overall pattern is quite similar in all four cases. In particular, for extremely low and high values of the Jacobi constant escaping motion completely dominates, while on the other hand, close encounter motion is more prominent for intermediate values of $C$. It should be noted, that the energy range for which close encounter motion has non-zero values decreases as the central primary becomes more massive. Indeed, when $\beta = 0.05$ this interval is roughly about $-2 < C < 4$, while for $\beta = 50$ it shrinks to $-1 < C < 1$. We may argue that these energy interval are the most important ones, where the orbital structure of the dynamical system is more rich and therefore interesting.

Our analysis suggests that mainly in the above-mentioned energy intervals the average escape as well as the close encounter times of the orbits display a common hierarchy. In particular, as the value of the mass parameter $\beta$ increases (which implies a more massive central body, with respect to the other three equally massed peripheral bodies) both the average escape and close encounter times of orbits seem to increase.

So far, we discussed the degree of fractality only by using qualitative arguments, regarding the geometry of the $(x,C)$ plane. However it would very informative to present some quantitative arguments, regarding the parametric evolution of the degree of fractality. A convenient way is by computing the uncertainty or fractal dimension, by following the approach used in \cite{AVS01,AVS09}. In panel (a) of Fig. \ref{frac} we depict the evolution of the uncertainty dimension $D_0$, as a function of the mass parameter $\beta$. It is evident that the fractal dimension reduces, as the mass of the central primary body increases, following an exponential decay.

Very recently, a new quantitative method, for measuring the degree of the basin fractality (or unpredictability), was introduced \cite{DWGGS16}. This new tool, which is called ``basin entropy" $S_b$, examines the topological properties of the basin diagrams and provides accurate quantitative results about their degree of fractality. The numerical algorithm of how basin entropy works is explained in detail in \cite{DWGGS16}.

In panel (b) of Fig. \ref{frac} we present the parametric evolution of the basin entropy of the $(x,C)$ plane, as a function of the mass parameter $\beta$. Obviously, for creating this diagram we used information not only from the four cases shown in Fig. \ref{xC}, but also from additional cases. It is observed that the basin entropy exhibits a smooth decrease, with increasing value of $\beta$. This phenomenon however, was anticipated because as we seen in Fig. \ref{xC}(a-b) the area on the $(x,C)$ plane covered by a complicated mixture of initial conditions of orbits (the regions in the exterior region between the two stability islands) decreases, as we proceed to higher values of the mass parameter. Nevertheless, we found that the reduction of the degree of fractality (as it was measured by both the uncertainty dimension and the basin entropy) of the $(x,C)$ plane follows a specific smooth and monotonic trend.

\section{The networks of periodic orbits}
\label{nets}

\subsection{Families of symmetric periodic orbits}
\label{sym}

In this subsection we will apply the grid method \cite{MBM74} for obtaining the network of the families of the symmetric periodic orbits of the system (see e.g., \cite{BP11}). Our numerical search will be expanded for periodic orbits with multiplicity $n$ up to ten. In other words, we will obtain those periodic orbits which display up to twenty cuts with the $x$-axis, per period. At this point, we would like to explain that the choice of the maximum value of the multiplicity $(n = 10)$ is not entirely arbitrary. In particular, we decided to search for periodic orbits with multiplicity up to ten in an attempt to obtain a complete view, regarding the network of the families of the periodic orbits of the system. Our numerical analysis takes place for four values of the mass parameter $\beta$, in order to determine the influence of this parameter on the network of the periodic orbits.

\begin{figure*}[!t]
\centering
\resizebox{\hsize}{!}{\includegraphics{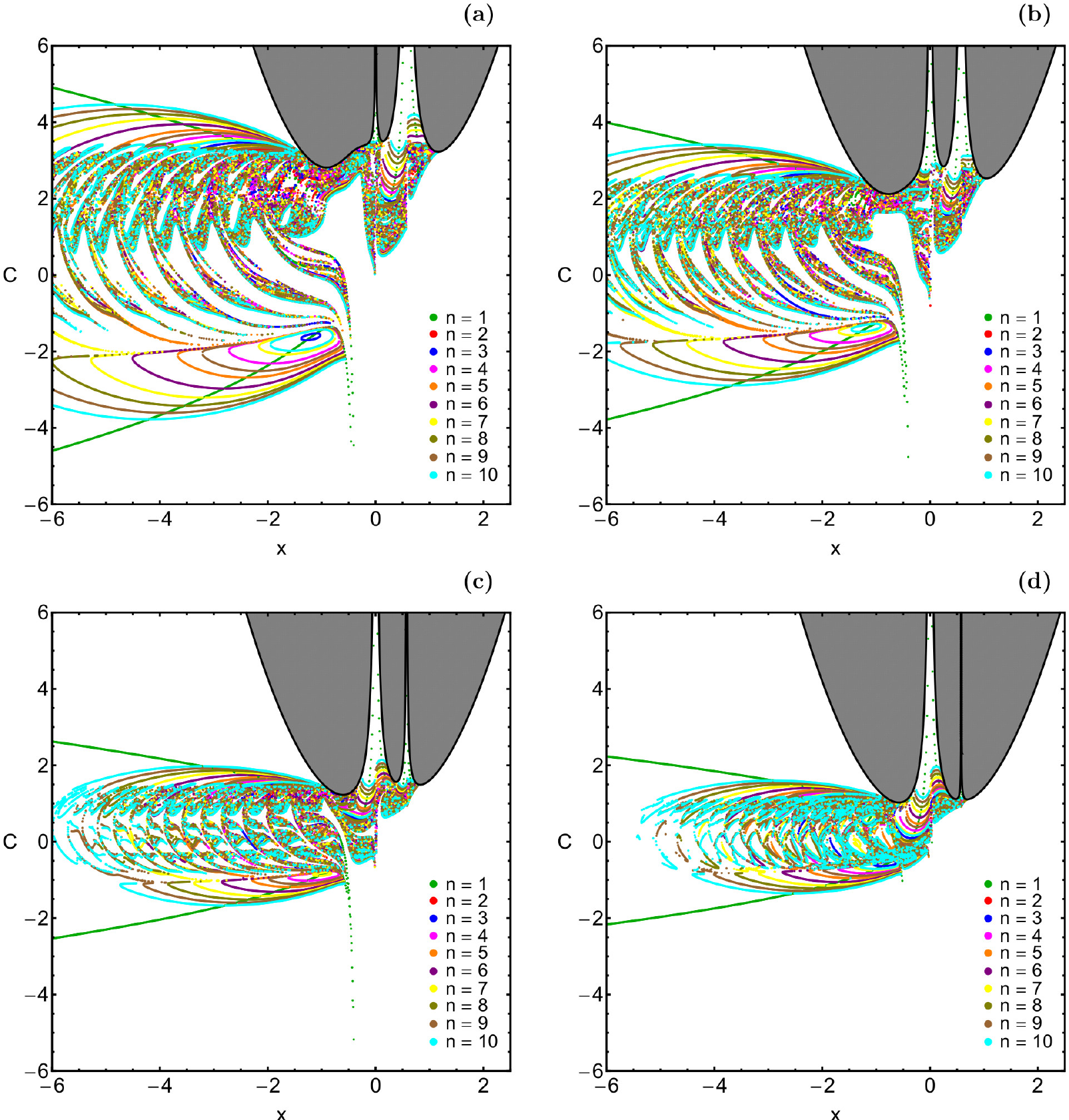}}
\caption{The network of the families of the symmetric periodic orbits, up to ten multiplicity, on the $(x,C)$ plane, for (a-upper left): $\beta = 0.05$; (b-upper right): $\beta = 0.5$; (c-lower left): $\beta = 5$; (d-lower right): $\beta = 50$. The energetically forbidden regions are shown in gray. (Color figure online).}
\label{fpo}
\end{figure*}

The presentation of the network of the periodic orbits, on the $(x,C)$ place, takes place through those initial conditions which lead to close periodic orbits, after 1, 2, ..., 10 oscillations. These initial conditions form the characteristic curves of the families of the symmetric periodic orbits. It should be noted that we present those periodic orbits, with initial conditions $x_0$, $y_0 = \dot{x_0} = 0$, and $\dot{y_0} > 0$ (i.e. ``positive" perpendicular intersection with the $x$-axis). In Fig. \ref{fpo}(a-d) we illustrate the network of the families of the symmetric periodic orbits (with $n = 1, 2, ..., 10$), for four values of the mass parameter $\beta$. The several colors represents the available values of the multiplicity $n$. For instance, with green color we depict the simple periodic orbits with multiplicity $n = 1$ (those periodic orbits which display two perpendicular cuts with the $x$-axis, per period), with red color the symmetric periodic orbits with multiplicity $n = 2$, etc.

Looking at the four panels of Fig. \ref{fpo} it becomes evident that the mass parameter $\beta$ strongly influences the network of the families of the symmetric periodic orbits. This is true because for relatively small values of $\beta$ the periodic solutions expand in a vast region of the $(x,C)$ plane. With increasing value of the mass parameter however, the area of the periodic solutions is reduced and confined. When $\beta = 0.05$ the vast majority of the periodic solution appear in the energy range $-6 < C < 6$, while the energy range is confined to $-2 < C < 2$, for $\beta = 50$. Additional numerical calculations for $\beta > 50$ indicate that the overall structure of the network of the symmetric periodic orbits on the $(x,C)$ plane remains almost unperturbed, by the shift on the value of the mass parameter.

Another interesting conclusion of the study of the network of the families of the symmetric periodic orbits is the following: it seems that for high enough values of the mass parameter $\beta$ the families of the simple periodic orbits (those with only two vertical cuts on the $x$-axis, per period) define the lower as well as the upper limits on the $(x,C)$ plane, in which all the periodic orbits of higher multiplicity exist and evolve. For example, for $\beta = 0.05$ (see panel (a) of Fig. \ref{fpo}) it is seen that the lower characteristic curve of the simple periodic orbits (with $n = 1$) is intersected by all the closed curves, which correspond to periodic orbits of higher multiplicity. In addition, these closed curves expand below the characteristic curve of the simple periodic orbits. On the other hand, for $\beta = 50$ (see panel (d) of Fig. \ref{fpo}), all the closed curves of the symmetric periodic orbits with $n > 1$ exist and evolve entirely between the two characteristic curves, formed by initial conditions of simple periodic orbits.

Therefore, we may conclude that when the mass of the central primary body is significantly larger than the masses of the other three peripheral primaries, the two characteristic curves of the simple periodic orbits define the region on the $(x,C)$ plane, where all the symmetric periodic orbits of higher multiplicity $(n > 1)$ exist and evolve.

\subsubsection{Stability of periodic solutions}
\label{stb}

\begin{figure*}[!t]
\centering
\resizebox{\hsize}{!}{\includegraphics{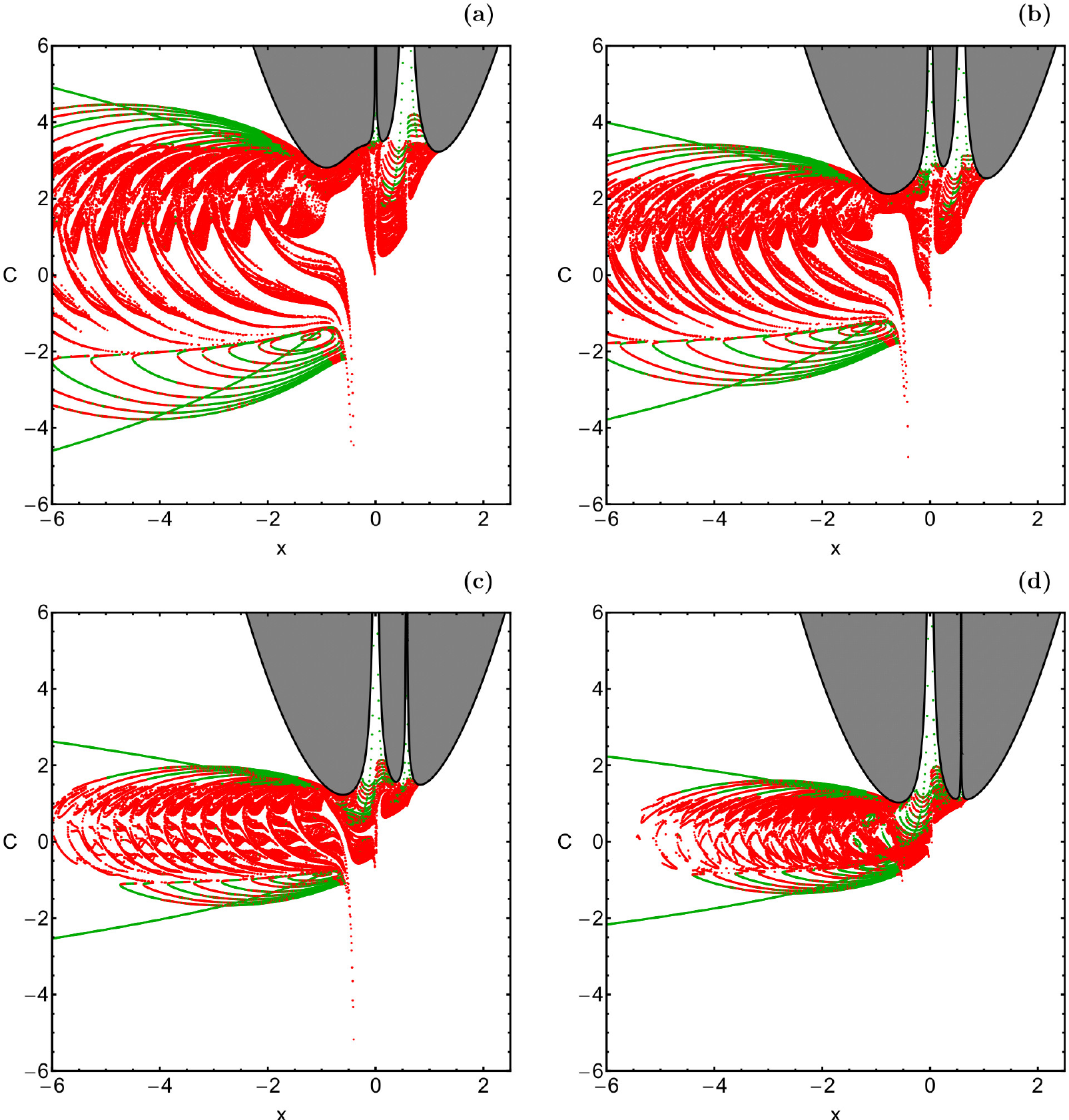}}
\caption{The linear stability (green) and non-stability (red) arcs of the families of the symmetric periodic orbits, on the $(x,C)$ plane, for (a-upper left): $\beta = 0.05$; (b-upper right): $\beta = 0.5$; (c-lower left): $\beta = 5$; (d-lower right): $\beta = 50$. The energetically forbidden regions are shown in gray. (Color figure online).}
\label{stab}
\end{figure*}

In the previous subsection, we numerically computed the initial conditions of the symmetric periodic orbits with multiplicity up to ten, for four values of the mass parameter $\beta$. One of the most important properties of the periodic solutions is their linear stability, which characterizes not only the periodic orbits themselves but also the dynamical system in general. For this purpose, we also studied the stability of all the symmetric families of periodic orbits (in other words, for all multiplicities) and for all values of the mass parameter.

The horizontal as well as the vertical iso-energetic linear (first approximation) stability of the symmetric periodic orbits is determined by computing the corresponding stability parameters $a_h$ and $a_v$, as they have been defined in \cite{H65}. For obtaining the linear stability of a periodic orbits we need to know its initial conditions with accuracy. Therefore, we numerically calculated the exact initial conditions of the periodic orbits as follows: when the initial conditions of a periodic orbit are derived, by the grid method, with accuracy according to the step size of the grid, then we proceeded to the calculation of the exact initial conditions of the periodic solution, using the standard corrector procedure, by numerically integrating the equations of motion along with the variational equations. The desired accuracy of the numerical integration in this work corresponds to twelve significant decimal figures, while the numerical error regarding the initial conditions of a periodic orbit is less than $10^{-8}$. As an error in periodicity we define the sum of the absolute difference between the initial numerical values (for $t = 0$) and the final numerical values (for $t = T$), for all variables. In all periodic orbits computed in this work, the numerical value of the Jacobi constant is conserved, during the numerical integration, with accuracy regarding the first 8 significant decimal digits. Especially in the case of the critical periodic solutions of the problem the accuracy is significantly increased regarding the first 15 significant decimal digits of $C$.

The linear stability of the symmetric periodic orbits with multiplicity up to ten, for the four values of the mass parameter $\beta$, is given in Fig. \ref{stab}(a-d). One can observe that the vast majority of the symmetric periodic orbits are unstable, while they are mainly located in the central region of the $(x,C)$ plane. Furthermore, stable periodic solution appear mainly at the lower and upper part of the region, defined by the periodic initial conditions. Stable periodic orbits are also observed between the energetically forbidden regions of motion. This fact is very interesting because there exist the families of the simple periodic solutions. Therefore, it is seen that in the vicinity of the simple periodic orbits there exist families composed mainly of stable periodic orbits. It is also of particular interest that the mass parameter $\beta$ does not practically affect the overall structure of the families of the periodic orbits as well as the ratio between stable and unstable periodic solutions.

The above-mentioned important properties of the simple symmetric periodic solutions of the dynamical system lead us to further numerically investigate them. Our numerical outcomes are presented in the following subsection.

It should be noted here that in the next subsection, we will encounter families which contain periodic solutions which tend to collide with one of the primaries. The relations (\ref{dist}) define the distances between the position of the test particle and the centers of the primary bodies. According to the bibliography, a collision with one of the primaries at $t^{*}$ is defined when the distance between the test particle and the respective primary tends to zero, when $t \to t^{*}$. In this work, when we state that a particular family leads to collision with one of the primaries, we simply imply that the distance between the test particle and the respective primary is smaller with respect to the minimum distance of the previous solutions, while it reduces for all next periodic solutions of the same family. Practically, we stop the computation of an orbital family when a periodic solution is obtained with a minimum distance between the test particle and the primary which is less that $10^{-5}$.

\subsubsection{Symmetric periodic solution}
\label{sps}

\begin{figure*}[!t]
\centering
\resizebox{\hsize}{!}{\includegraphics{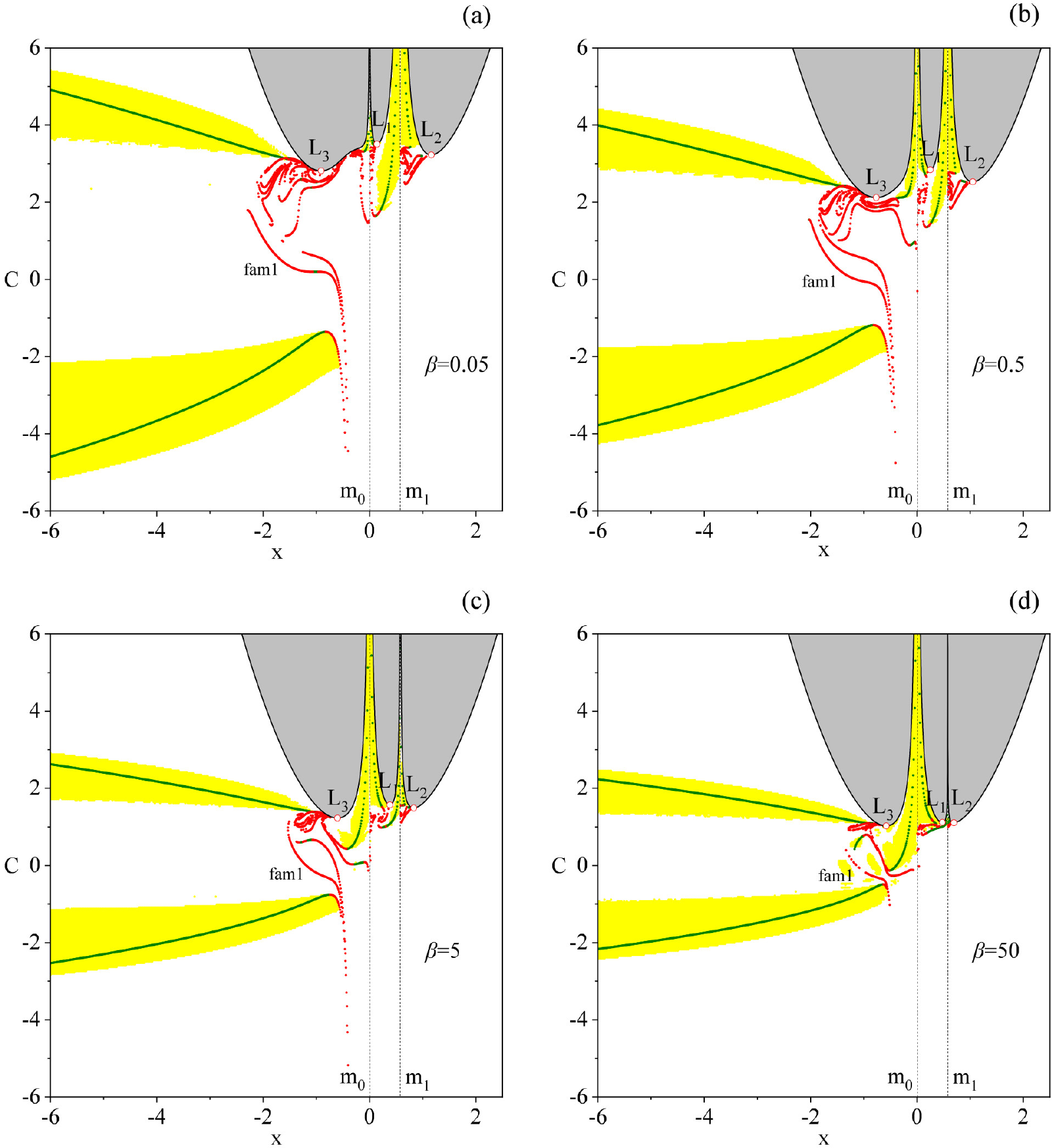}}
\caption{The network of simple $(n = 1)$ stable (green) and unstable (red) symmetric periodic orbits combined with the basins of non-escaping regular motion (yellow). Small red circles denote the positions of the collinear equilibrium points of the problem, while the vertical, dashed lines indicate the centers of the two primary bodies $P_0$ and $P_1$. The energetically forbidden regions are shown in gray. (Color figure online).}
\label{grd}
\end{figure*}

Studying the families of the simple periodic solutions of the problem we present in this subsection the network of all the respective families on the $(x,C)$ plane. Moreover, we choose one particular family of the simple periodic solutions, hereafter fam1, and we explore not only the shape and form of the corresponding orbits but also their time evolution. In addition, for every simple periodic solution we compute both the horizontal and the vertical stability indices. Our aim is to determine the influence of the mass parameter $\beta$ on the dynamical properties of this family of periodic orbits.

In Fig. \ref{grd}(a-d) we see how the initial conditions of stable (green) and unstable (red) simple $(n = 1)$ periodic orbits are distributed on the $(x,C)$ plane, for four values of the mass parameter $\beta$. In the same figure we also include the basins of non-escaping regular motion, shown earlier in Fig. \ref{xC}(a-d). We observe the excellent agreement of the results, that is how the stable branches of the simple periodic orbits coincide with the simultaneous presence of islands of quasi-periodic orbits. On the contrary, in the areas where the branches of the unstable periodic orbits exist there is no
numerical indication of quasi-periodic stability islands. Furthermore, it is evident, that when the mass of the central primary body is significantly larger than the masses of the other three primaries the number of the families is reduced, while at the same time the area on the $(x,C)$ plane, occupied by these symmetric periodic orbits, shrinks. In Fig. \ref{grd} the positions of the collinear equilibrium points $L_1$, $L_2$, and $L_3$ are pinpointed by red circles, while the black, vertical, dashed lines indicate the centers of the two primary bodies $P_0$ and $P_1$.

By computing the linear stability of the periodic solutions of family fam1 we also simultaneously derived both the horizontal and vertical-critical periodic solutions. The critical periodic solutions of a family are of paramount importance since from the horizontal-critical ones new families of coplanar periodic orbits are bifurcated, while the vertical-critical solutions are starting points for the determination of new families of three-dimensional periodic orbits. A periodic orbit is horizontal-critical when $|a_h| = 1$, while is vertical-critical when $|a_v| = 1$ (more details are given in \cite{H73}). A periodic solution is horizontal stable if $|a_h| < 1$, and vertical stable if $|a_v| < 1$. The vertical-critical periodic orbits are also known as \emph{generating planar orbits} because, as we have already mentioned, they are used as initial conditions for locating new families of periodic orbits of the system in three dimensions.

\begin{figure*}[!t]
\centering
\resizebox{\hsize}{!}{\includegraphics{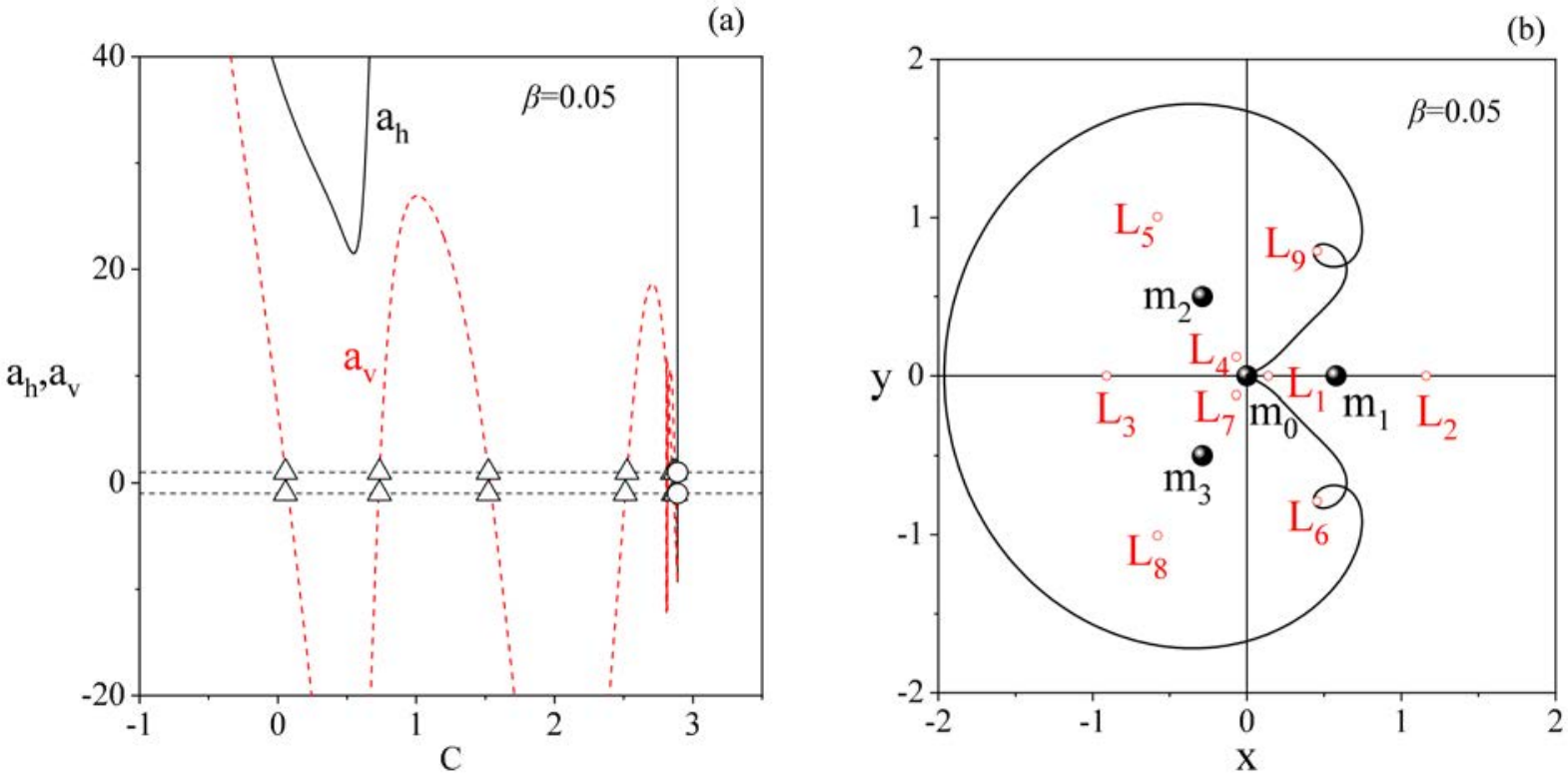}}
\caption{(a-left): The linear stability diagram of the family fam1 for $\beta = 0.05$. The black line is the curve of the horizontal stability parameter $a_h$, while the red, dashed line is the curve of the vertical stability parameter $a_v$, respectively. Small circles denote the horizontal-critical symmetric periodic orbits, while small triangles the vertical-critical ones. (b-right): An asymptotic symmetric periodic orbit, of the family fam1, circulating around the equilibrium points $L_6$ and $L_9$. (Color figure online).}
\label{cas1}
\end{figure*}

\begin{figure*}[!t]
\centering
\resizebox{\hsize}{!}{\includegraphics{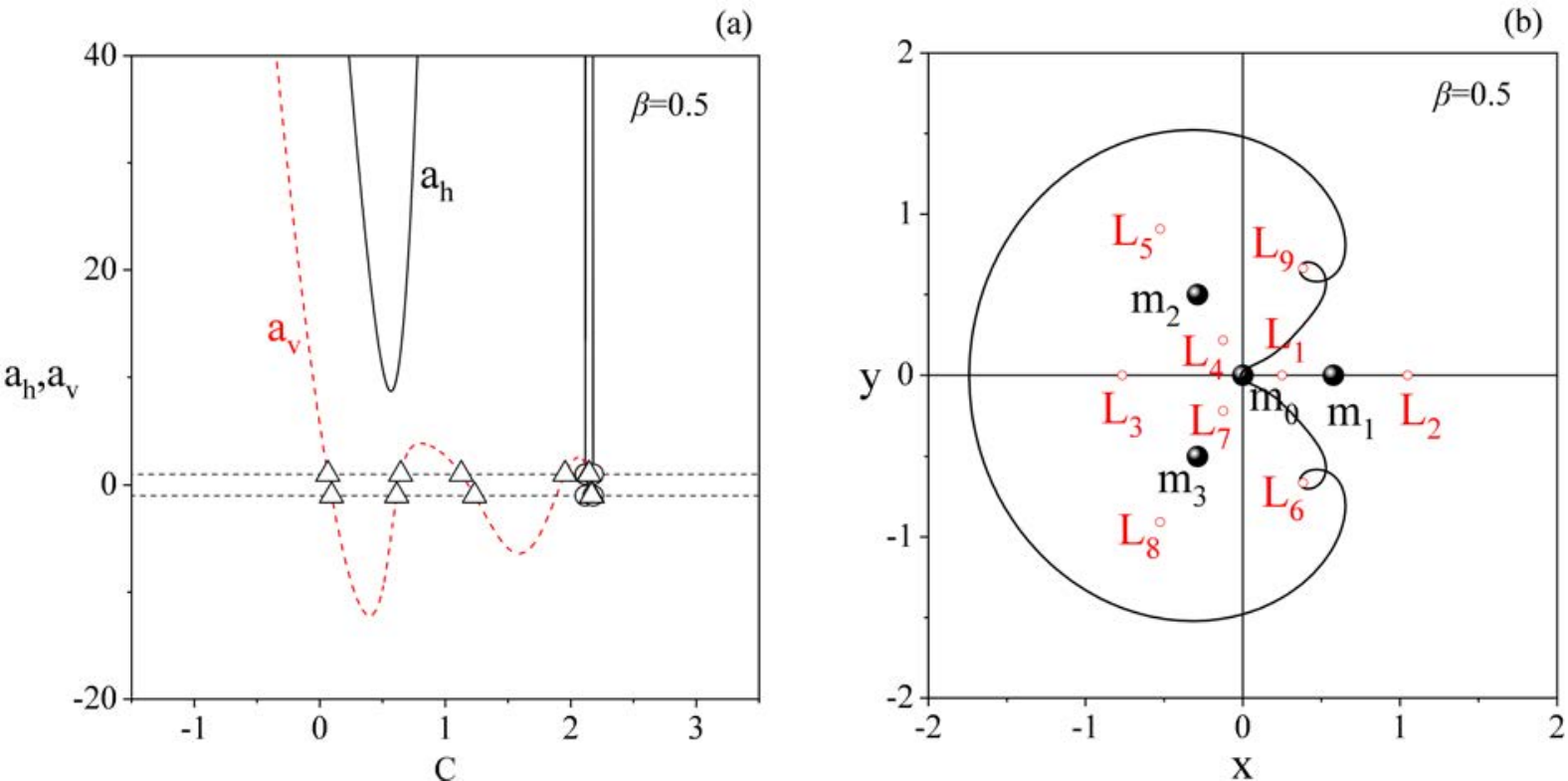}}
\caption{(a-left): The linear stability diagram of the family fam1 for $\beta = 0.5$. The color code is the same, as in Fig. \ref{cas1}. (b-right): An asymptotic symmetric periodic orbit, of the family fam1, circulating around the equilibrium points $L_6$ and $L_9$. (Color figure online).}
\label{cas2}
\end{figure*}

When $\beta = 0.05$ the family fam1 is composed of simple periodic orbits which retrogradely circulate around the primary bodies $P_2$ and $P_3$, while the pass very close to the central primary $P_0$. With decreasing time the periodic orbits of family fam1 gradually grow, while when the value of the Jacobi constant $C$ becomes lower than $C = -6$ (which has been set as the lower limit) the numerical search of periodic orbits is terminated. On the other hand, with increasing time the family evolves and the corresponding periodic orbits tend to become asymptotic to the libration points $L_6$ and $L_9$, as it is seen in panel (b) of Fig. \ref{cas1}. In the same panel figure, the positions of the centers of the primary bodies as well as of the equilibrium points are also indicated.

In panel (a) of Fig. \ref{cas1} we provide the linear stability diagrams, regarding family fam1, when $\beta = 0.05$. It is observed that the family fam1 is mainly composed of unstable periodic orbits. However as the time increases, and the family tends to obtain asymptotic solutions, stable solution appear. In the same diagram the vertical-critical periodic solutions are indicated by small triangles, while for the locations of the horizontal-critical solutions we use small circles. The two horizontal, dashed lines correspond to $a_h = a_v = \pm 1$ and delimit the region inside which stable periodic orbits exist. In Table \ref{tab1} we indicatively provide the initial conditions of some of the critical periodic solutions of the family fam1. Specifically, we provide the initial coordinate $x_0$ (the initial condition on the $x$-axis), the value of the Jacobi constant $C$, the time $t = T/2$ (where $T$ is the period of the orbits) and also the type of critically stability (horizontally or vertically) of the periodic orbits.

\begin{table*}[!t]
\caption[]{Horizontal or vertical-critical symmetric simple periodic orbits of the family fam1, for various values of the mass parameter $\beta$.}
\centering
\begin{tabular}{lcrrrl}
\hline\noalign{\smallskip}
\multicolumn{1}{c}{$\beta$} & \multicolumn{1}{c}{$x_{0}$} & \multicolumn{1}{c}{$x_1(T/2)$}
& \multicolumn{1}{c}{C} & \multicolumn{1}{c}{$T/2$} & \multicolumn{1}{c}{Stability} \\  \noalign{\smallskip}\hline\noalign{\smallskip}
0.05  & $-1.94706852$         & 1.36681371 &    2.89010610 & 6.90841642 & $a_h=+1$ \\
      & $-1.94625290$         & 1.36580561 &    2.89009006 & 6.92559247 & $a_h=-1$ \\
      & $-0.61359465$         & 1.71571146 &    0.05579570 & 0.62710591 & $a_v=+1$ \\
      & $-1.93318541$         & 1.35504692 &    2.87519719 & 7.49666798 & $a_v=-1$ \\
0.5   & $-1.73996371$         & 1.27856489 &    2.11507380 & 9.74521359 & $a_h=+1$ \\
      & $-1.72977479$         & 1.24144732 &    2.17748185 & 6.63993269 & $a_h=-1$ \\
      & $-0.63215646$         & 1.44615679 &    0.06411314 & 0.71873136 & $a_v=+1$ \\
      & $-1.72308172$         & 1.24526456 &    2.14768035 & 7.72896645 & $a_v=+1$ \\
5     & $-1.31834719$         & 1.24760350 &    0.60024370 & 3.22322701 & $a_h=+1$ \\
      & $-1.07971350$         & 1.00481407 &    0.66484805 & 2.33212070 & $a_h=+1$ \\
      & $-1.30407179$         & 1.23399371 &    0.60108926 & 3.16858364 & $a_v=+1$ \\
      & $-0.69292699$         & 1.02557354 &    0.19530569 & 1.06869814 & $a_v=-1$ \\
50    & $-1.01081257$         & 0.80795618 &    0.76750530 & 2.92223247 & $a_h=-1$ \\
      & $-0.96512775$         & 0.74391131 &    0.79535015 & 2.76389457 & $a_h=-1$ \\
      & $-1.16480271$         & 1.10691256 &    0.47763655 & 3.14395386 & $a_v=+1$ \\
      & $-0.90794526$         & 0.70066209 &    0.77676887 & 2.40747815 & $a_v=-1$ \\
\noalign{\smallskip}\hline
\end{tabular}
\label{tab1}
\end{table*}

For $\beta = 0.5$ the structure of the family fam1 qualitatively does not change, in relation to what we seen for $\beta = 0.05$. The shape of the corresponding periodic orbits remains the same, while the pattern of their evolution has the same characteristics: namely, from one side the family fam1 goes outside of the limit $C = -6$ (and we stopped to calculate the rest of the family), and from the other side the family tends to, as the time $t \to \infty$, in asymptotic orbits around the critical equilibrium points $L_6$ and $L_9$ (see panel (b) of Fig. \ref{cas2}). When $\beta = 0.5$ the family fam1 contains both horizontal and vertical-critical periodic solutions (see panel (a) of Fig. \ref{cas2}) , while in Table \ref{tab1} we provide the details of some of them.

\begin{figure}[!t]
\centering
\resizebox{0.7\hsize}{!}{\includegraphics{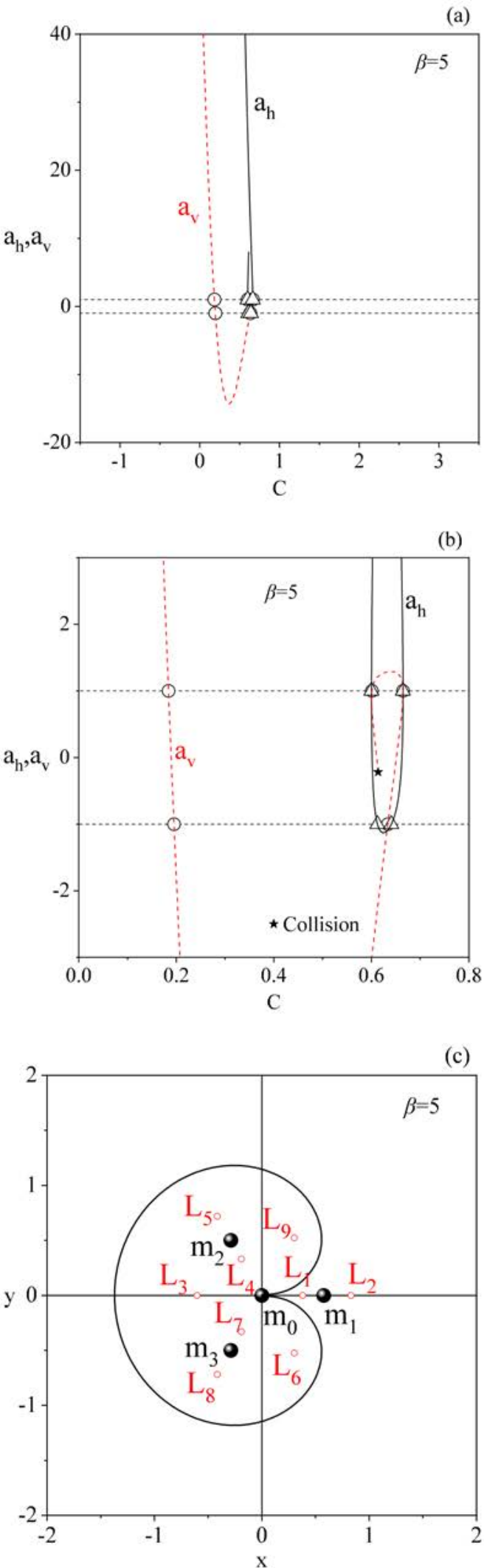}}
\caption{(a-upper panel): The linear stability diagram of the family fam1 for $\beta = 5$. The color code is the same, as in Fig. \ref{cas1}. (b-middle panel): Magnification of the area close to the critical values of the stability parameters. (c-lower panel): A symmetric periodic orbit, of the family fam1, just before the collision of the test particle with the central primary body $P_0$. (Color figure online).}
\label{cas3}
\end{figure}

\begin{figure}[!t]
\centering
\resizebox{0.6\hsize}{!}{\includegraphics{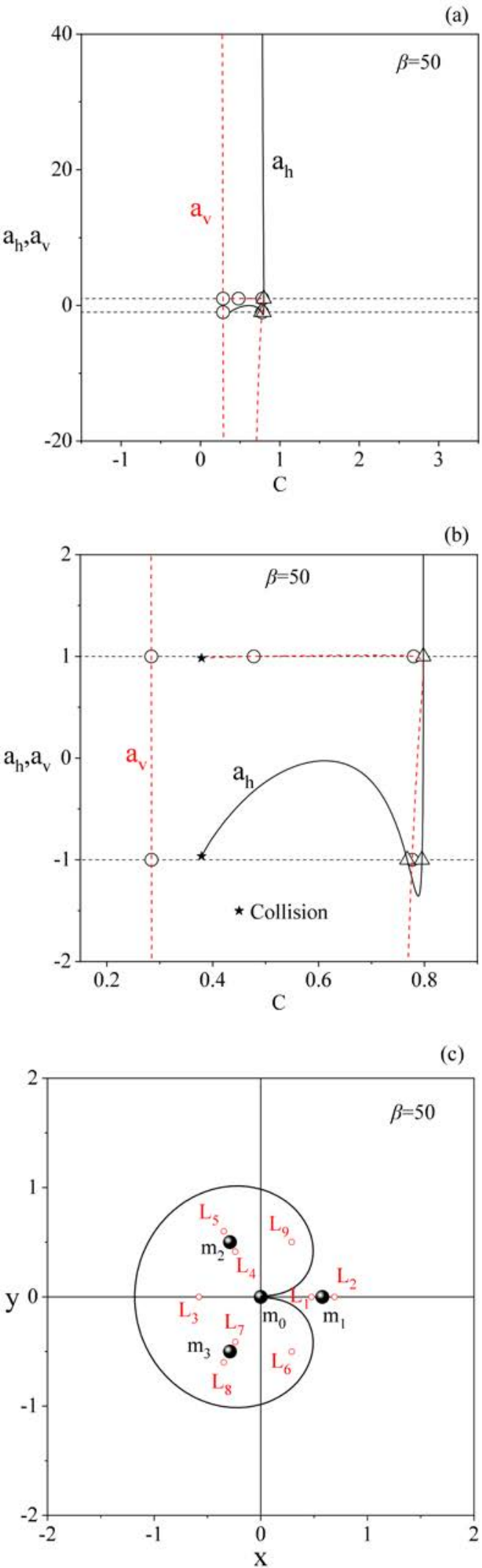}}
\caption{(a-upper panel): The linear stability diagram of the family fam1 for $\beta = 50$. The color code is the same, as in Fig. \ref{cas1}. (b-middle panel): Magnification of the area close to the critical values of the stability parameters. (c-lower panel): A symmetric periodic orbit, of the family fam1, just before the collision of the test particle with the central primary body $P_0$. (Color figure online).}
\label{cas4}
\end{figure}

\begin{figure}[!t]
\centering
\resizebox{0.6\hsize}{!}{\includegraphics{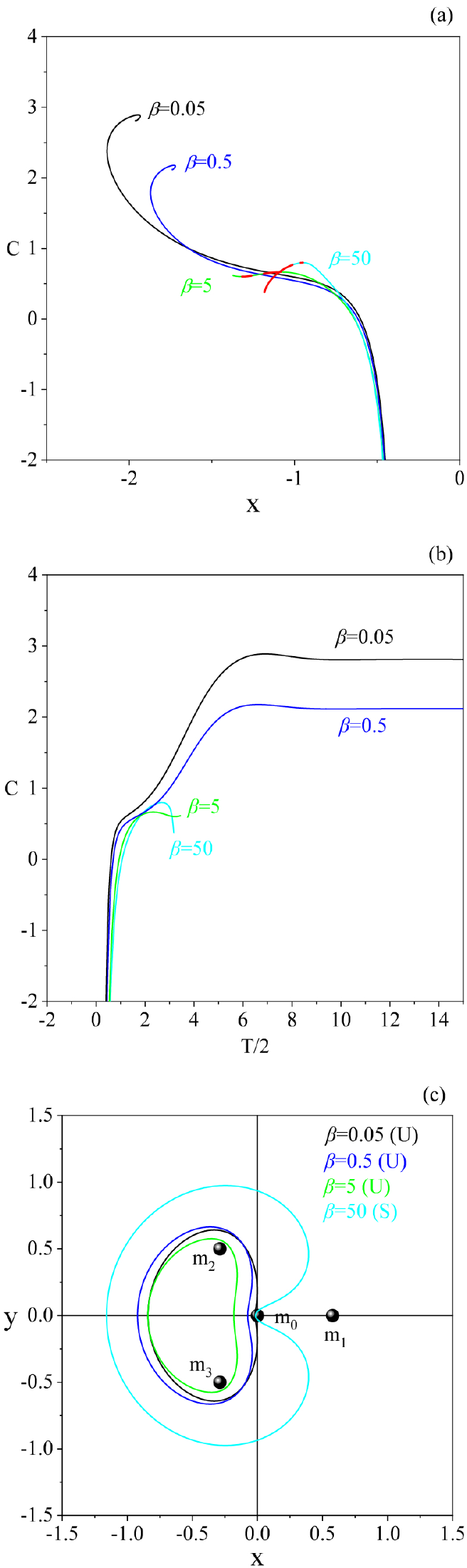}}
\caption{(a-upper panel): Characteristic curves of the family fam1 for $\beta = \{0.05, 0.5, 5, 50\}$. The horizontal linear stability arcs of the families are also shown in red. (b-middle panel): Evolution of the half-period time $T/2$ of the family fam1, for the four cases, regarding the value of the mass parameter $\beta$. (c-lower panel): Four simple symmetric periodic orbits for $\beta = \{0.05, 0.5, 5, 50\}$, for the same value of the Jacobi constant $C = 0.5$. The capital letter U denote unstable periodic orbits, while the capital letter S indicate stable periodic orbits. (Color figure online).}
\label{char}
\end{figure}

For the next two cases, that is for $\beta = 5$ and $\beta = 50$, the content of the periodic network changes since the amount of stable periodic orbits increases (see panels (a-b) of Figs. \ref{cas3} and \ref{cas4}). In panels (b) of Figs. \ref{cas3} and \ref{cas4} we present local magnifications of panels (a), near the stability regions $-1 < a_h, a_v < 1$, where we have a clear picture, regarding the horizontal and vertical-critical periodic orbits, for $\beta = 5$ and $\beta = 50$. An additional important different between the first two ($\beta = 0.05$ and $\beta = 0.5$) and the last two cases ($\beta = 5$ and $\beta = 50$) is that in the last two cases the families of the periodic orbits do not lead to asymptotic solution but to collision with the central primary body $P_0$. In panels (b) of Figs. \ref{cas3} and \ref{cas4} a five-pointed star indicates the collision of the periodic orbits, for the two cases of the mass parameter $\beta$. Moreover, in panels (c) of Figs. \ref{cas3} and \ref{cas4} we present two periodic orbit, just before collision occur for family fam1, for $\beta = 5$ and $\beta = 50$, respectively. Some characteristic examples of critical periodic orbits of the family fam1, for the last two cases of the mass parameter, are also given in Table \ref{tab1}.

For better comparison, in panel (a) of Fig. \ref{char} we depict the characteristic curves on the $(x,C)$ plane, for all four families of periodic orbits and for all four values of the mass parameter $\beta$. In the same diagram, using red color, we also present the linear stability arcs. It should be noted, that the extremely small energy ranges of stable periodic orbits, for $\beta = 0.05$ and $\beta = 0.5$, are not visible. From this diagram it becomes evident that the mass parameter $\beta$ greatly influences the families of the periodic orbits. Indeed, with increasing value of the mass parameter the several families of periodic orbits shrink, while at the same time the energy regions of stable periodic solutions grow.

The evolution of the half-period $T/2$, for all four cases, is given in panel (b) of Fig. \ref{char}. It is observed, that for the cases with $\beta = 0.05$ and $\beta = 0.5$, where asymptotic solutions exist, the time of the half-period tends to infinity. On the contrary, for the cases with $\beta = 5$ and $\beta = 50$ the evolution of the half-period time ends, when upon the collision of the test particle with the central primary body $P_0$.

Finally, in panel (c) of Fig. \ref{char} we illustrate, with different colors, the shape of four symmetric periodic orbits, for the four values of the mass parameter $\beta$. For being able to directly compare these four periodic orbits we chose, for all of them, the same value of the Jacobi constant $C = 0.5$. The linear stability of the periodic orbits is also given in the same diagram, where it is seen that first three of them (corresponding to $\beta = \{0.05, 0.5, 5\}$) are unstable, while the periodic orbit corresponding to $\beta = 50$ is stable.

\subsection{Families of non-symmetric periodic orbits}
\label{asymf}

One of the most important issues in dynamical systems is the study of non-symmetric periodic orbits and particularly of simple non-symmetric periodic orbits, with only two intersections with the horizontal $x$-axis, per period. At this point, it should be noted that when we talk about symmetric and non-symmetric periodic orbits we refer to symmetry of the periodic solution, with respect to the horizontal $x$-axis of the system.

The determination of the families of the symmetric periodic solutions is feasible by using the grid method, as it was explained in the previous subsection \ref{sym}, and by obtaining the set $(x_0, y_0 = 0, \dot{x_0} = 0, \dot{y_0}(x_0,C_0) > 0)$. On the other hand, the task of locating non-symmetric periodic orbits is by far more complicated, since now the required initial conditions are $(x_0, y_0 = 0, \dot{x_0}, \dot{y_0}(x_0,\dot{x_0},C_0) > 0)$. Therefore, the non-zero parameter $\dot{x_0}$ adds an extra dimension to the grid method. For this purpose, in many cases for locating non-symmetric periodic solutions, we first locate the families of the symmetric periodic solutions and then we compute the horizontal-critical periodic orbits with the properties $a_h = 1$ and $b_h = 0$ (see e.g., \cite{P08}). Then we are sure that this critical periodic solutions intersects a family of non-symmetric periodic orbits (for more details the reader is referred to \cite{H65,H05}). In other cases, depending of course on the particular symmetry of the system, we exploit the possible symmetry of the periodic solutions, with respect to the vertical $y$-axis. Then the set of the initial conditions is $(x_0 = 0, y_0, \dot{x_0}(y_0,C_0), \dot{y_0} = 0$, and therefore the grid method is also applicable for finding periodic solutions (if they exist) which are symmetric, with respect to the $y$-axis, but non-symmetric, with respect to the $x$-axis.

Due to the fact that the peripheral primary bodies have equal masses three axes of symmetries exist in the system under consideration. This means that we can use the presence of symmetries for searching for non-symmetric periodic solutions which are launched vertically from the axes of symmetry. We will also search for simple (with only two intersections with the $x$-axis, per period) non-symmetric periodic orbits, with constant $\dot{x_0} = const. \neq 0$, when the use of the grid method is applicable.

\begin{figure}[!t]
\centering
\resizebox{\hsize}{!}{\includegraphics{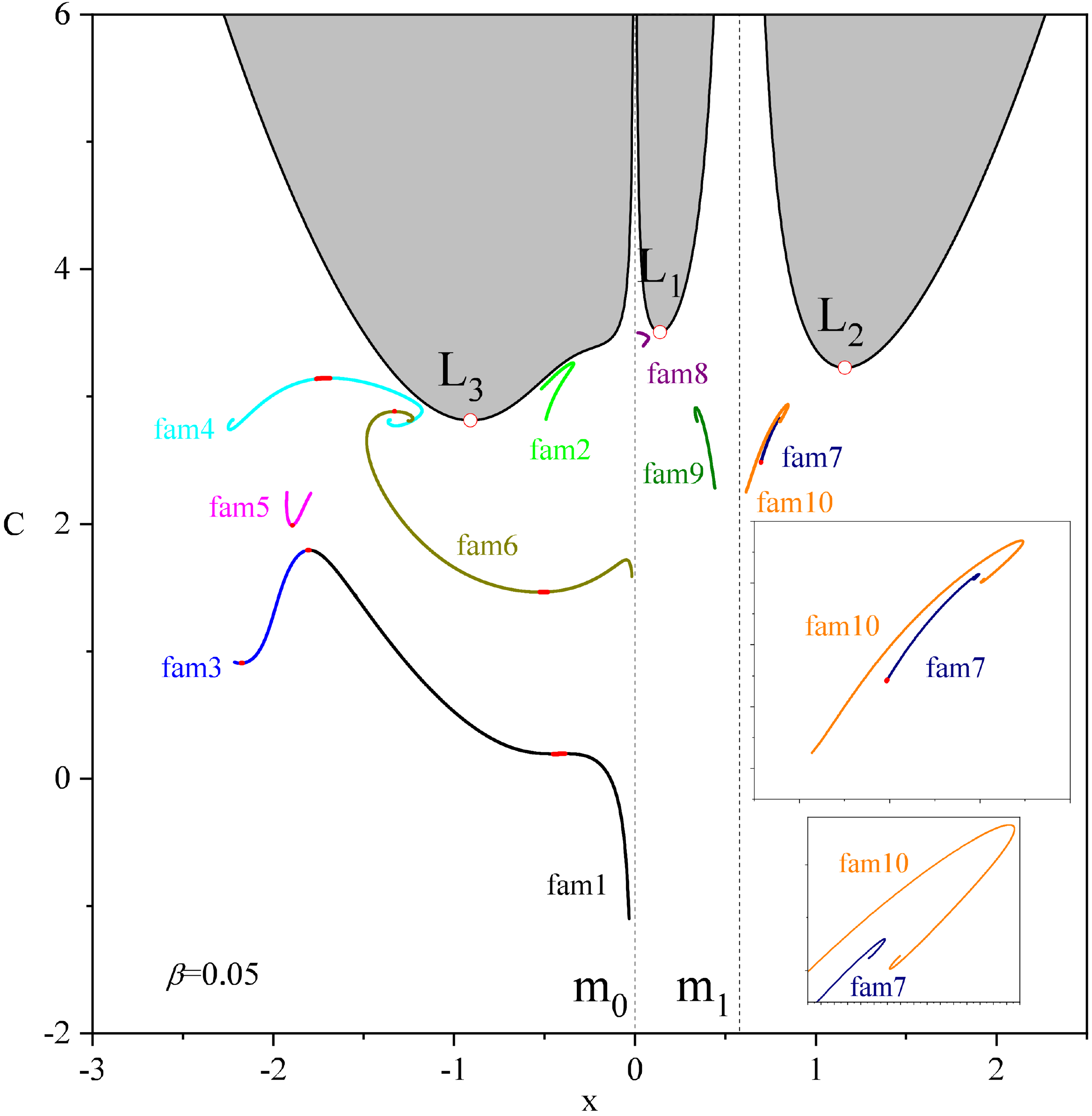}}
\caption{Characteristic curves of the ten families of non-symmetric periodic orbits of the problem for $\beta = 0.05$. The horizontal linear stability arcs of the families are also shown (red segments). Small, red circles denote the positions of the collinear equilibrium points, while the vertical, dashed lines indicate the position of the centers of the two primary bodies $P_0$ and $P_1$. The gray regions correspond to the energetically forbidden regions of motion. Each color corresponds to a different family of non-symmetric periodic orbits. Additional details are given using two frames containing in magnification the area close to the characteristic curves of the families 7 and 10. (Color figure online).}
\label{asym}
\end{figure}

\begin{figure}[!t]
\centering
\resizebox{0.7\hsize}{!}{\includegraphics{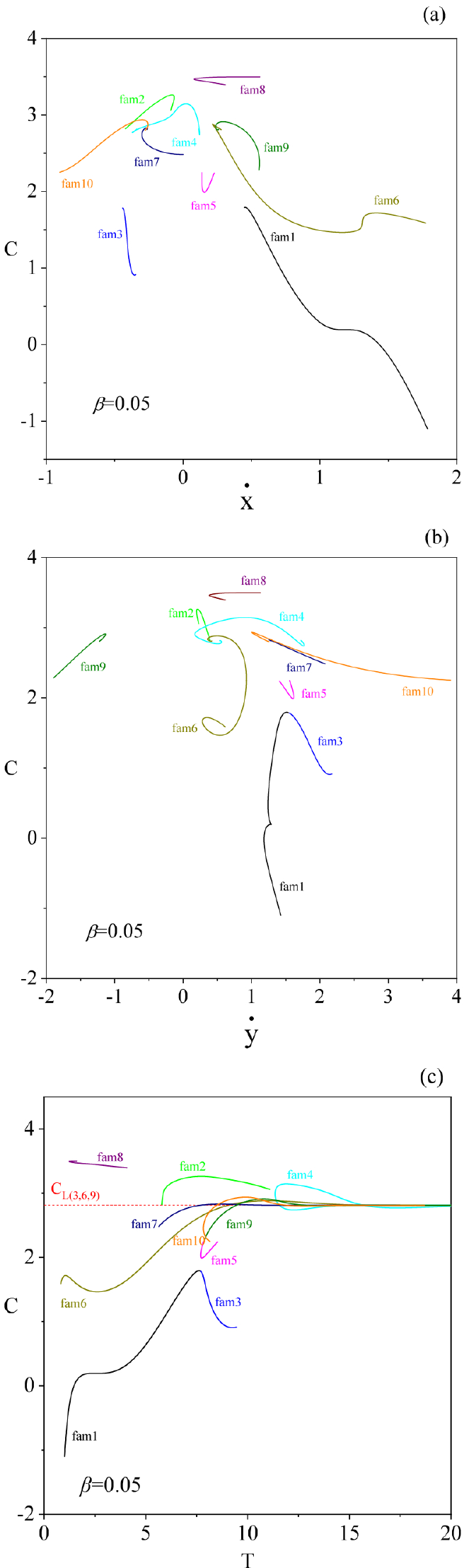}}
\caption{Characteristic curves of the ten families of non-symmetric periodic orbits of the problem for $\beta = 0.05$ which show the components of the velocity and the period $T$ of the periodic solutions. (Color figure online).}
\label{asym2}
\end{figure}

For every family of non-symmetric periodic solutions and also for every periodic orbit we will also compute the corresponding linear stability, in an attempt to present a complete view, regarding the network of the periodic orbits in the planar circular restricted five-body problem. For the numerical study of the horizontal-stability we use the stability index $S_h$ (see \cite{H65,H73}), which is defined as $S_h = \left(a_h + d_h \right)/2$, while for the vertical-stability we similarly use the index $S_v = \left(a_v + d_v \right)/2$. Specifically, the equations of motion (\ref{eqmot}) can also be written as
\begin{equation}
\dot{x}_i = f_i(x_1,\dots,x_6), \ \ \ i = 1,\dots,6,
\end{equation}
for $x_1 = x, x_2 = y, x_3 = z$, while
\begin{eqnarray}
&&f_1 = x_4 = \dot{x},
  f_2 = x_5 = \dot{y},
  f_3 = x_6 = \dot{z}, \nonumber \\
&&f_i = \frac{\partial\Omega}{\partial{x_i}}, \ \ \ i = 4,5,6.
\end{eqnarray}
If we define an iso-energetic mapping of the Poincar\'{e} surface of section $(x_1,x_4)$ into itself then the mapping can be described by
\begin{align}
x_1 &= F_1(x_{01}, x_{04}, C), \nonumber\\
x_4 &= F_2(x_{01}, x_{04}, C).
\end{align}
Then, the parameters of the horizontal linear stability are
\begin{equation}
a_h = \frac{\partial{F_1}}{\partial x_1}, \ \
b_h = \frac{\partial{F_1}}{\partial x_4}, \ \
c_h = \frac{\partial{F_2}}{\partial x_1}, \ \
d_h = \frac{\partial{F_2}}{\partial x_4},
\end{equation}
and they describe the linear stability of the planar periodic orbits under perturbations in the plane of motion, while they are also the coefficients from the variational matrix
\begin{eqnarray}
  \left(
           \begin{array}{c}
             \Delta x \\
             \Delta \dot{x} \\
           \end{array}
  \right)=\left(
           \begin{array}{cc}
             a_h & b_h \\
             c_h & d_h \\
           \end{array}
  \right)\left(
           \begin{array}{c}
             \Delta x_0 \\
             \Delta \dot{x}_0 \\
           \end{array}
  \right).
\end{eqnarray}

In the same vein, the parameters of the vertical stability are given by the following relations
\begin{equation}
a_v = v_{33}(T), \ \
b_v = v_{36}(T), \ \
c_v = v_{63}(T), \ \
d_v = v_{66}(T),
\end{equation}
when $v_{ij} = \partial x_i/\partial x_{0j}$, $i,j = 1,\dots,6$ and $T$ the full period time.

A non-symmetric periodic orbit is horizontal-stable if $-1 < S_h < 1$ and vertical-stable if $-1 < S_v < 1$, while the case of a critical solution appears when $S_h = \pm 1$ (horizontal-critical) and $S_v = \pm 1$ (vertical-critical). In the following subsections, we will also compute and present the critical periodic solutions of all the families of the periodic orbits.

\subsubsection{Non-symmetric periodic solutions for $\beta = 0.05$}
\label{nsp1}

Using the above-mentioned technique we managed to locate several families of simple non-symmetric periodic orbits, when $\beta = 0.05$. In this subsection, we are going to present in detail ten of the families of non-symmetric periodic solutions.

The characteristic curves of the ten families of simple non-symmetric periodic orbits on the $(x,C)$ plane, for $\beta = 0.05$, are illustrated in Fig. \ref{asym}. Here it should be emphasized that in the case of symmetric periodic solutions the data of the $(x,C)$ plane provide information for the full set of initial conditions $(x_0, y_0 = 0, \dot{x_0} = 0, \dot{y_0}(x_0,C_0))$ of the orbits. On the other hand, in the case of non-symmetric periodic solutions we cannot obtain information for the full set of initial conditions, mainly because the characteristic curves do not provide any kind of information, regarding the initial velocities of the test particle (fifth body). We chose these ten families of non-symmetric periodic orbits so as to cover almost all the available space on the $(x,C)$ plane, that is the left-hand side of $P_0$ (near the equilibrium point $L_3$), between the primaries $P_0$ and $P_1$ (near the equilibrium point $L_1$), as well as in the right-hand side of $P_1$ (near the equilibrium point $L_2$). For every non-symmetric periodic orbit we computed its linear stability and the corresponding results are also given in Fig. \ref{asym}, using red arcs.

\begin{figure*}[!t]
\centering
\resizebox{\hsize}{!}{\includegraphics{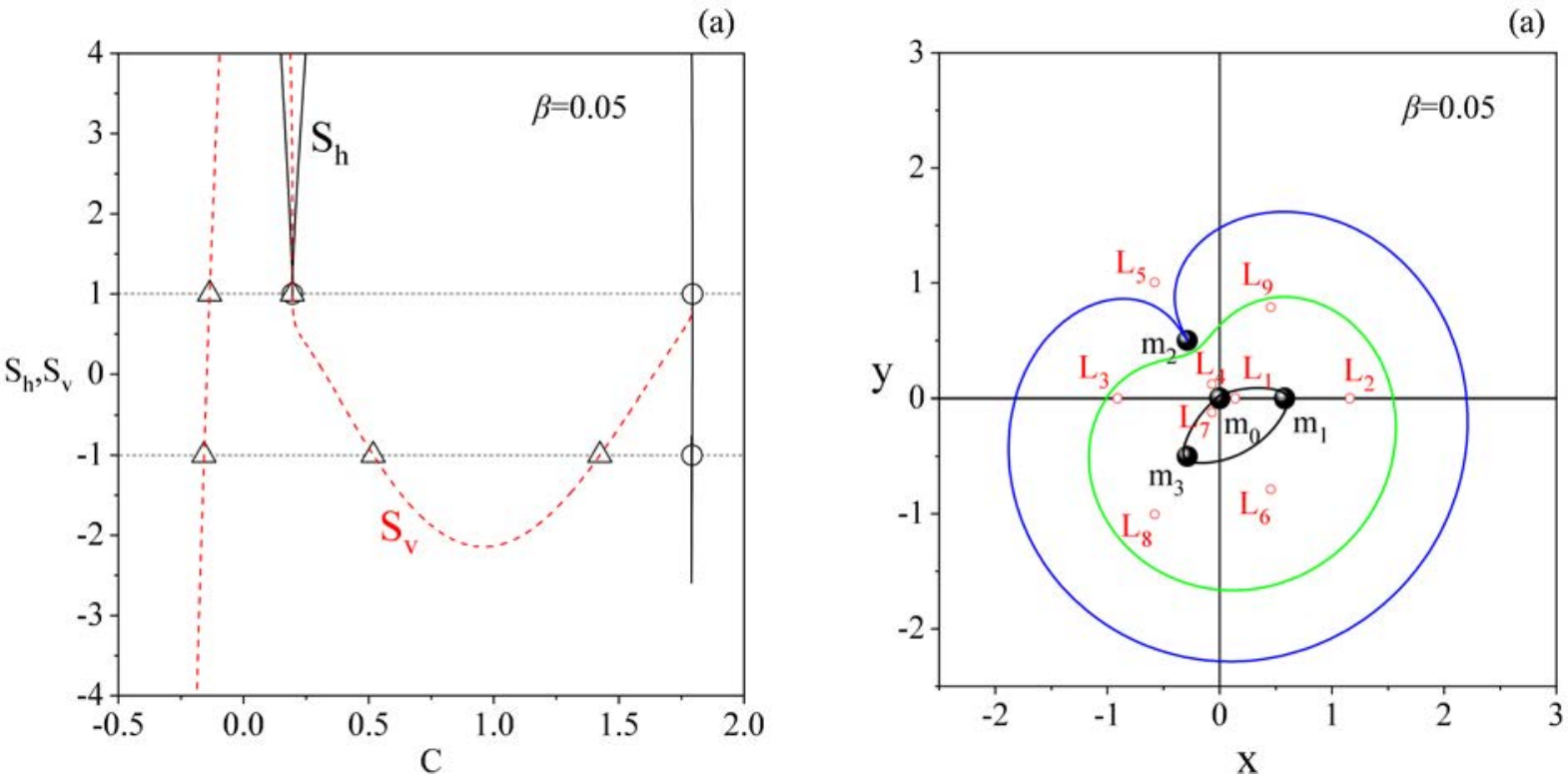}}
\caption{(a-left): Characteristic curves of the linear stability parameters $S_h$ (horizontal) and $S_v$ (vertical) of the family fam1, for $\beta = 0.05$. Small circles and triangles denote the critical-horizontal and vertical periodic orbits, correspondingly. (b-right): Three non-symmetric periodic orbits of this family. (Color figure online).}
\label{asymf1}
\end{figure*}

\begin{figure*}[!t]
\centering
\resizebox{\hsize}{!}{\includegraphics{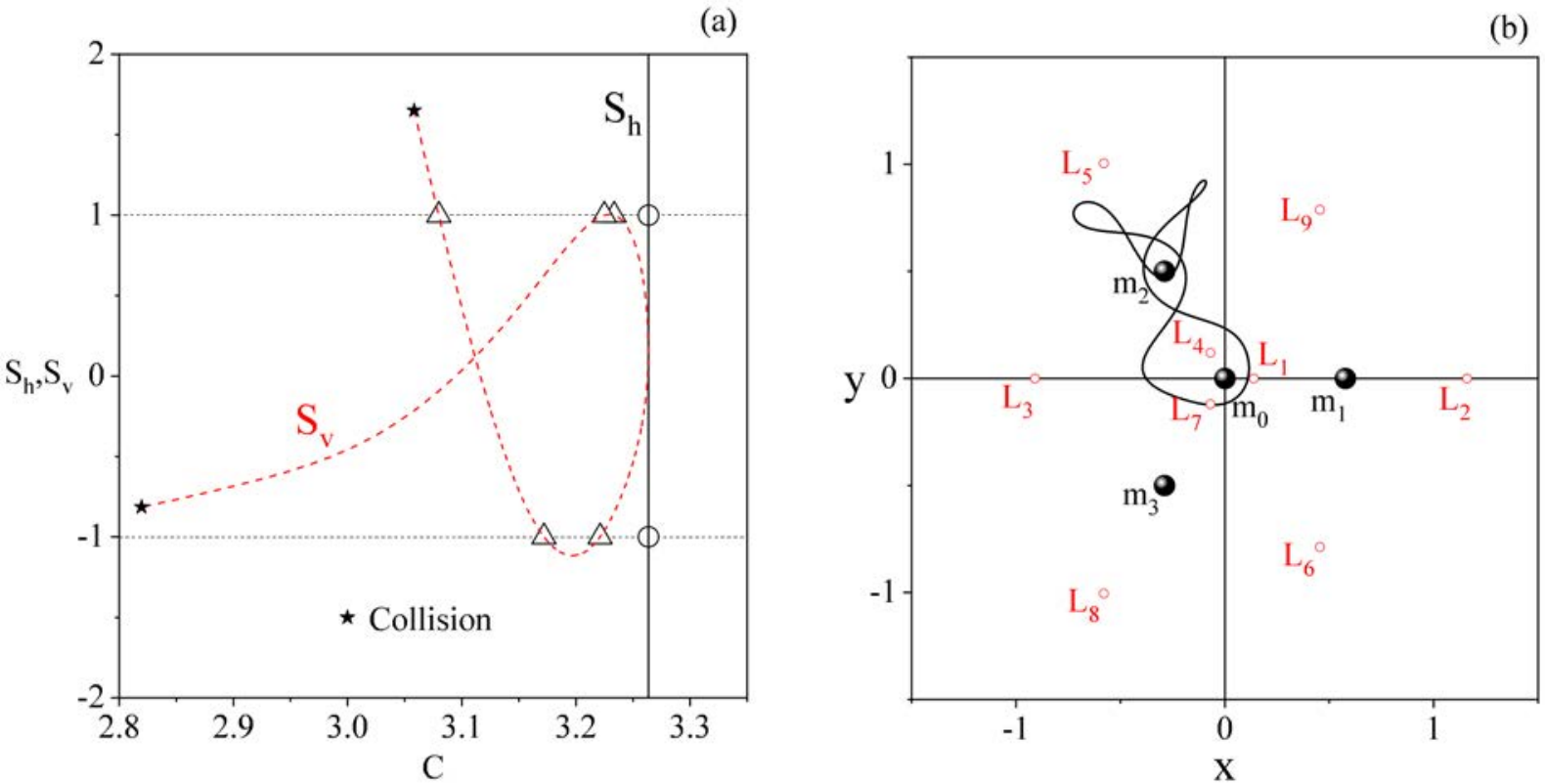}}
\caption{((a-left): The linear stability diagram of the family fam2, for $\beta = 0.05$. Stars denote the collision non-symmetric periodic orbits of the family. (b-right): A simple non-symmetric periodic orbit of this family. (Color figure online).}
\label{asymf2}
\end{figure*}

\begin{figure*}[!t]
\centering
\resizebox{\hsize}{!}{\includegraphics{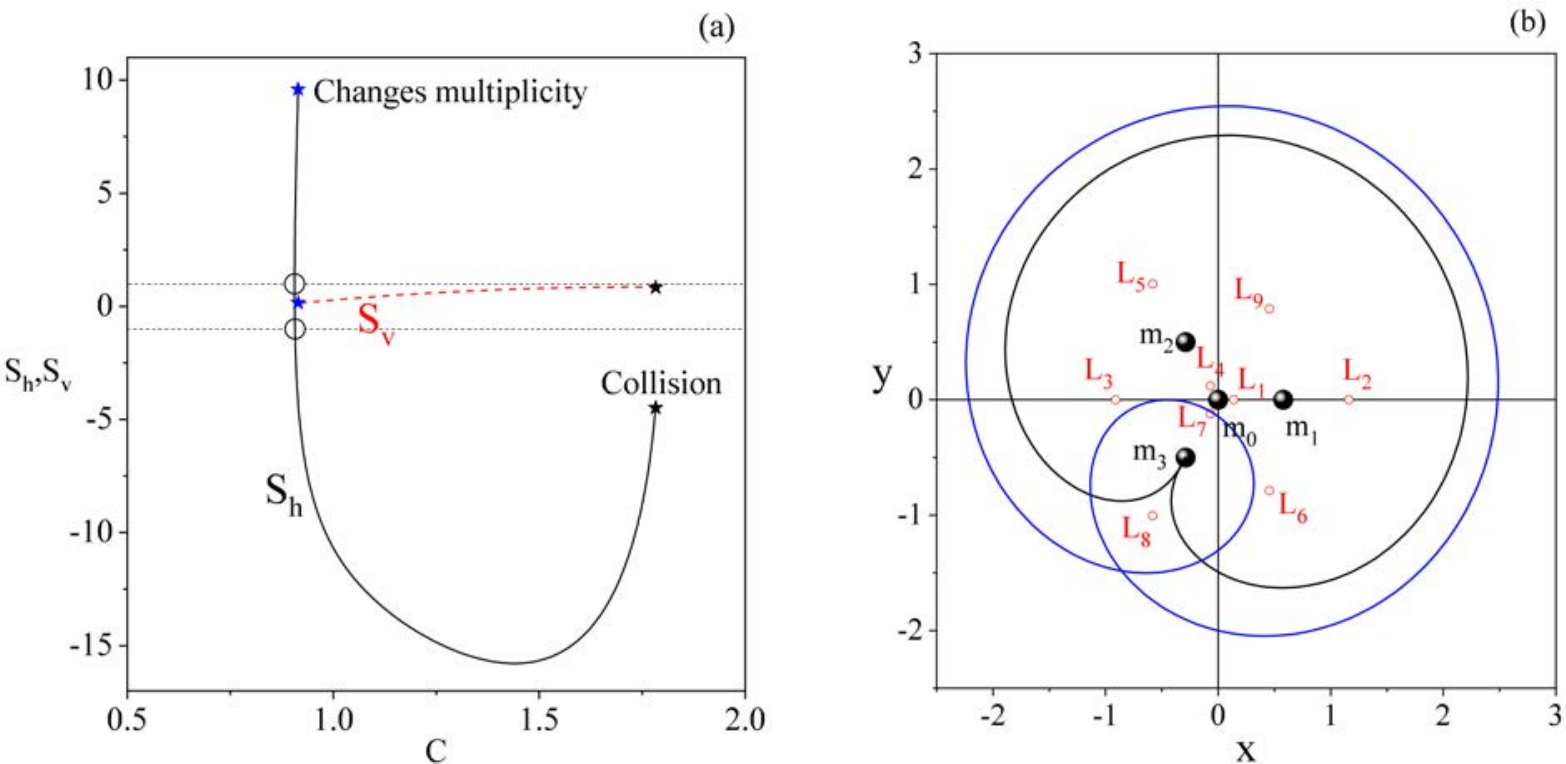}}
\caption{(a-left): The linear stability diagram of the family fam3, for $\beta = 0.05$. Black stars denote the collision non-symmetric periodic orbits, while blue ones the orbits which tend to change their multiplicity. (b-right): Two simple non-symmetric periodic orbits of this family. The black one tends to collision with the primary body $P_3$ and the second one (blue) tends to change its multiplicity. (Color figure online).}
\label{asymf3}
\end{figure*}

\begin{figure*}[!t]
\centering
\resizebox{\hsize}{!}{\includegraphics{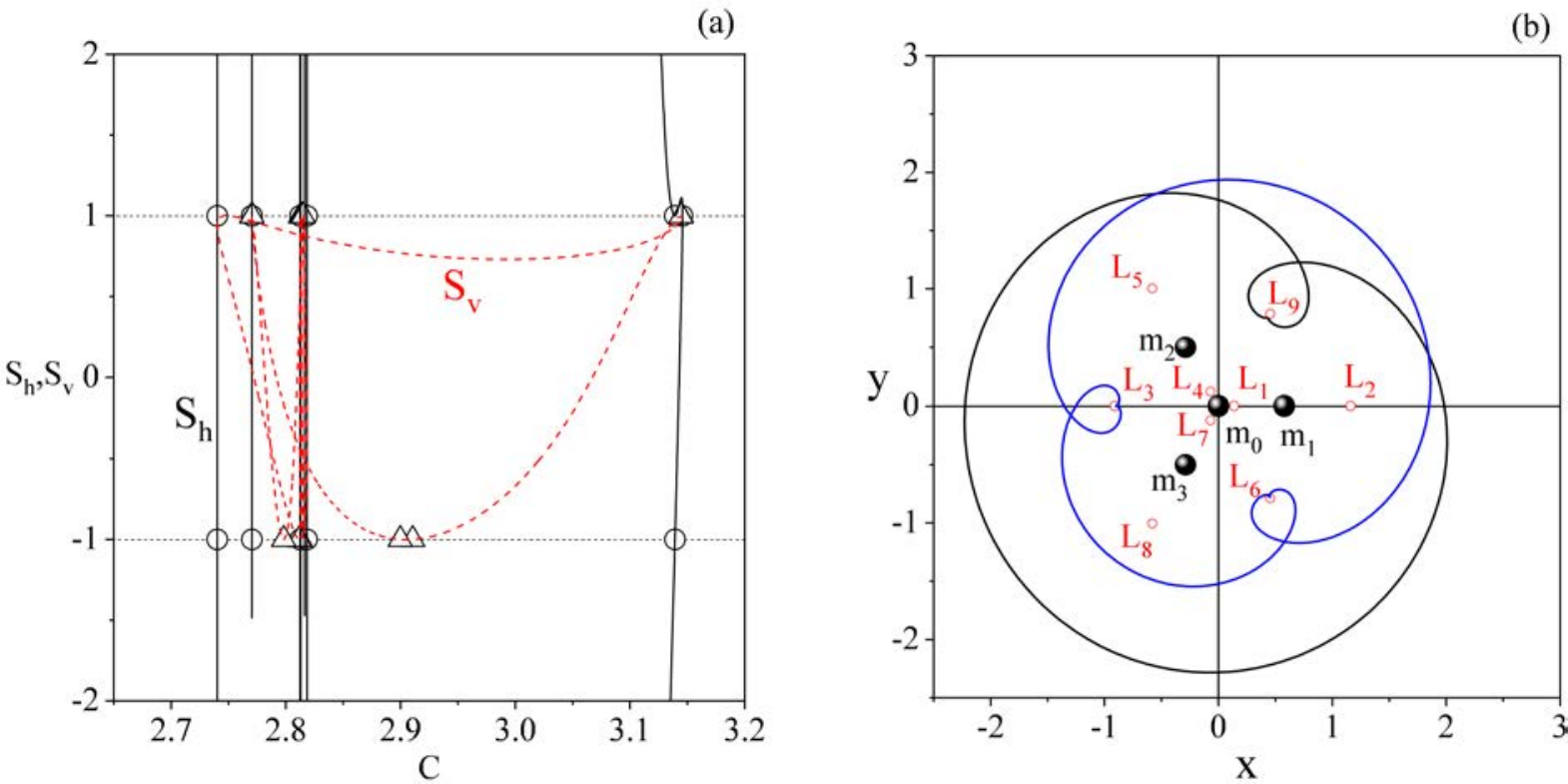}}
\caption{(a-left): The linear stability diagram of the family fam4, for $\beta = 0.05$. (b-right): Two non-symmetric periodic orbits of this family. The black one is an almost homoclinic asymptotic orbit around the equilibrium point $L_9$, while the blue one is an almost heteroclinic asymptotic orbit, around the equilibrium points $L_3$ and $L_6$. (Color figure online).}
\label{asymf4}
\end{figure*}

\begin{figure*}[!t]
\centering
\resizebox{\hsize}{!}{\includegraphics{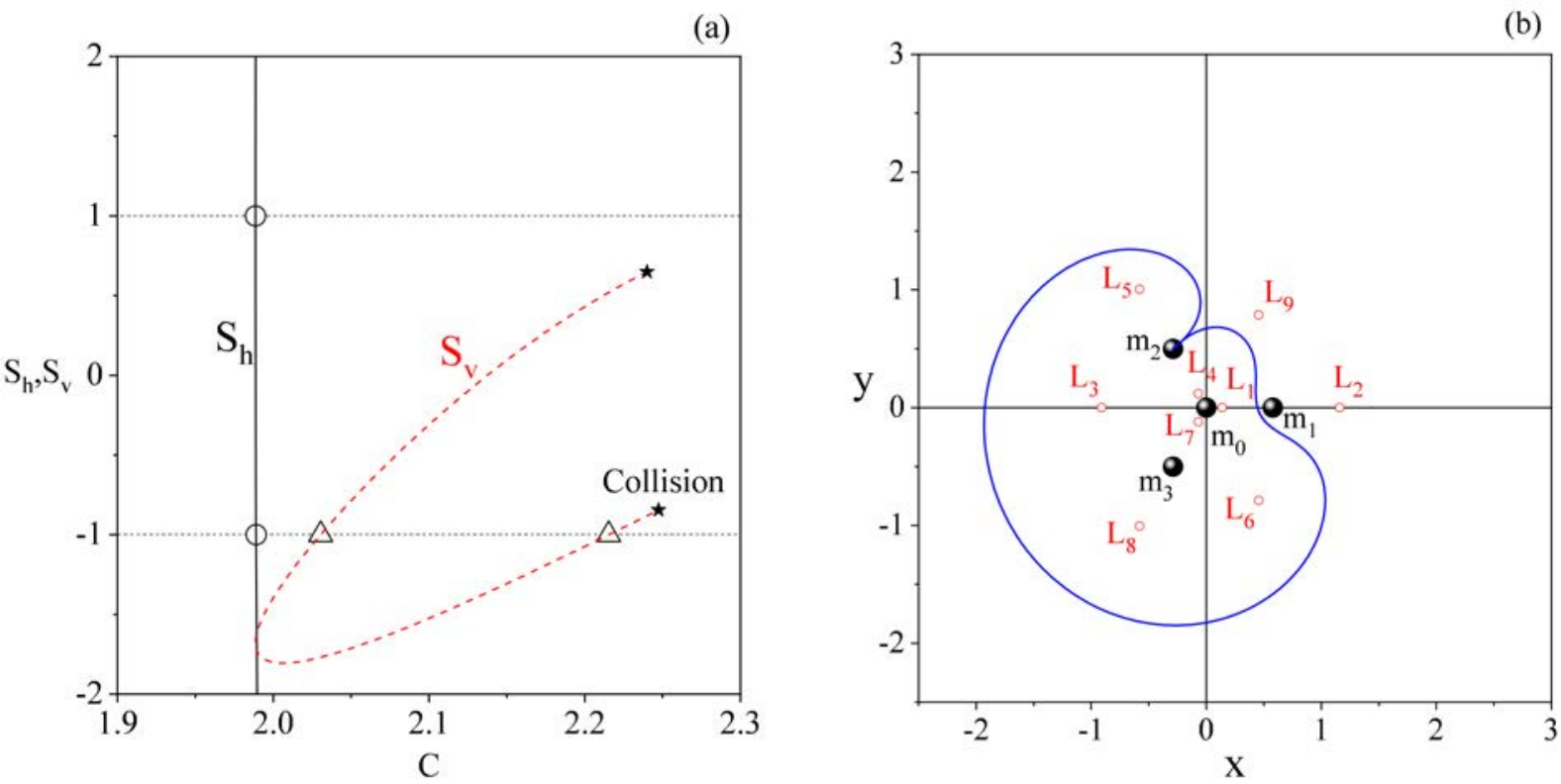}}
\caption{(a-left): The linear stability diagram of the family fam5, for $\beta = 0.05$. (b-right): A simple non-symmetric periodic orbit of this family just before the test particle collides with the primary body $P_2$. (Color figure online).}
\label{asymf5}
\end{figure*}

\begin{figure*}[!t]
\centering
\resizebox{\hsize}{!}{\includegraphics{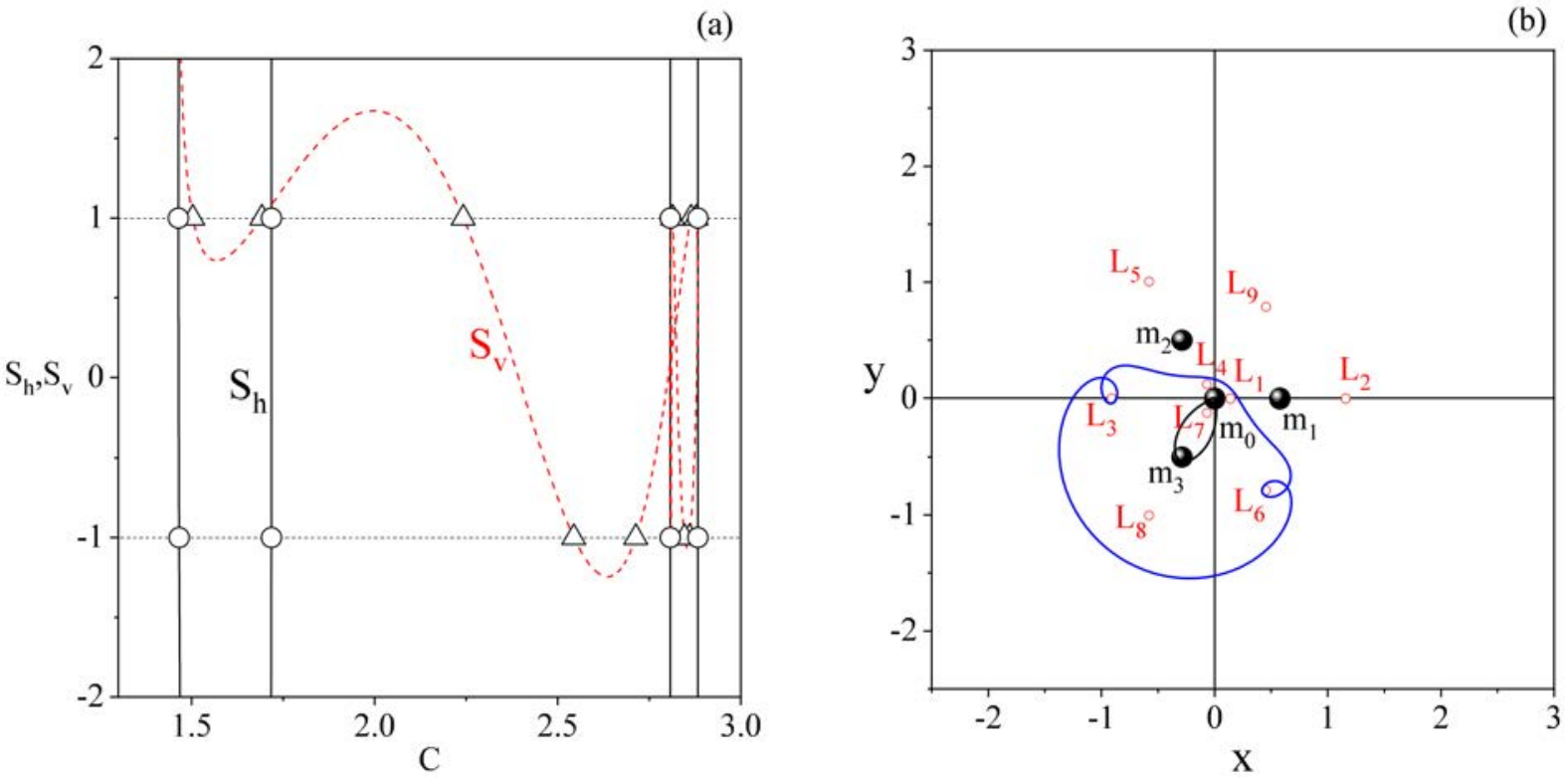}}
\caption{(a-left): The linear stability diagram of the family fam6, for $\beta = 0.05$. (b-right): Two non-symmetric periodic orbits of this family. The black one is an orbit just before the fifth body collides with the central primary body $P_0$, while the blue orbit is an almost heteroclinic asymptotic solution, around the equilibria $L_3$ and $L_6$. (Color figure online).}
\label{asymf6}
\end{figure*}

\begin{figure*}[!t]
\centering
\resizebox{\hsize}{!}{\includegraphics{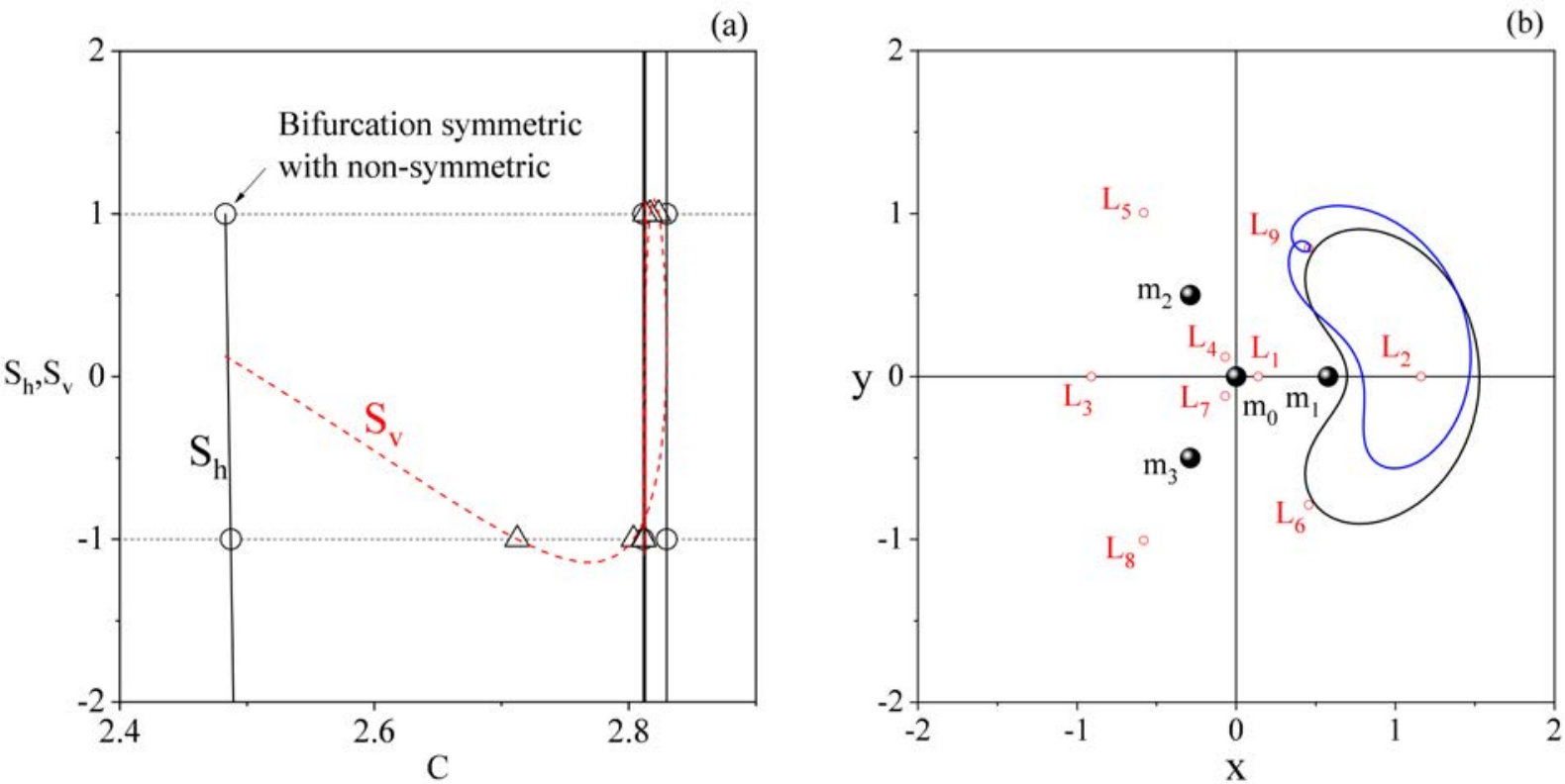}}
\caption{(a-left): The linear stability diagram of the family fam7, for $\beta = 0.05$. (b-right): Two periodic orbits of this family. The black one is a symmetric orbit which belongs in two families (a symmetric and a non-symmetric), while the blue orbit is an almost homoclinic
asymptotic solution around the equilibrium point $L_9$. (Color figure online).}
\label{asymf7}
\end{figure*}

\begin{figure*}[!t]
\centering
\resizebox{\hsize}{!}{\includegraphics{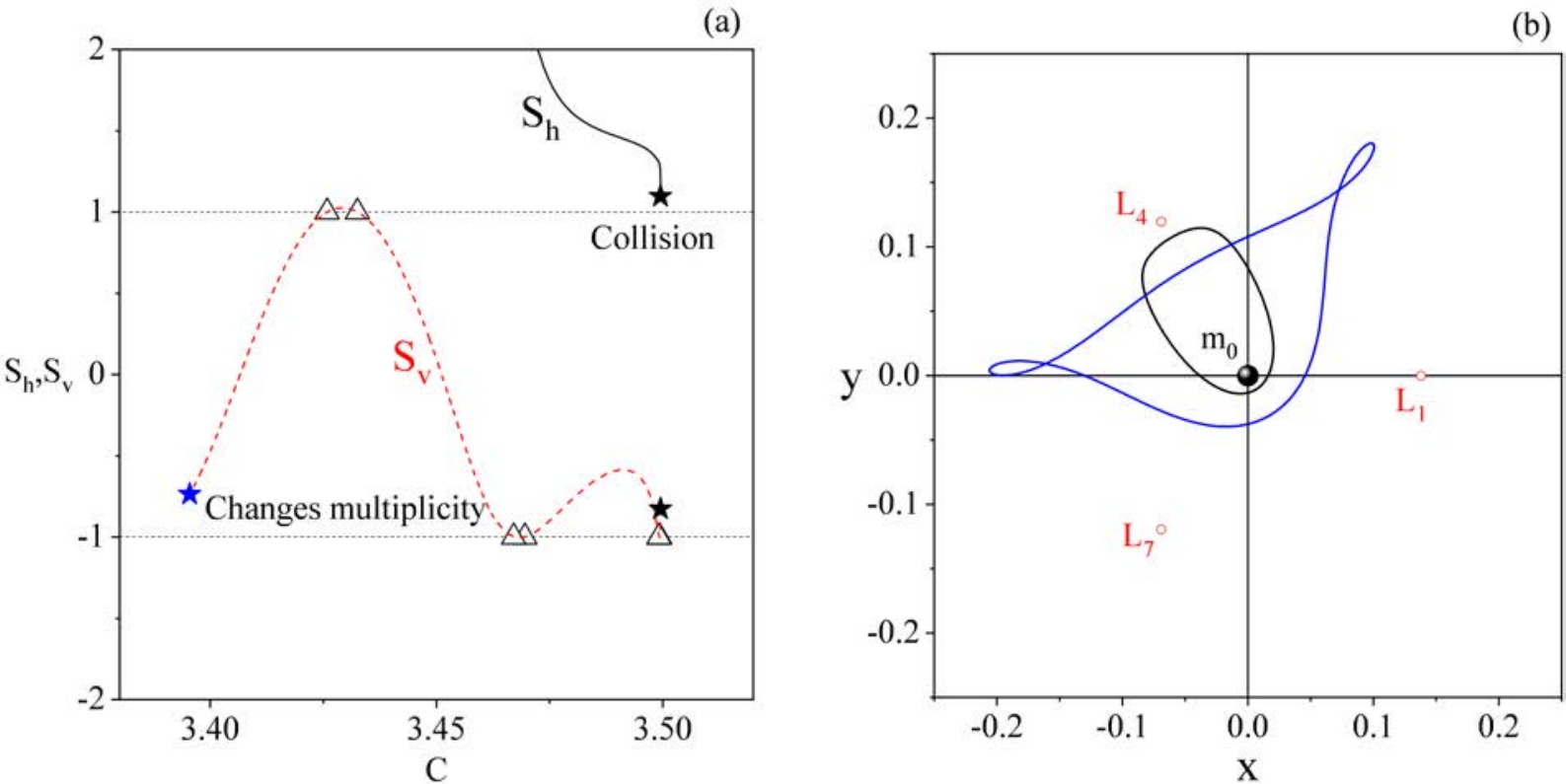}}
\caption{(a-left): The linear stability diagram of the family fam8, for $\beta = 0.05$. (b-right): Two periodic orbits of this family. The black one is just before the test particle collides with the central primary body $P_0$, while the blue orbit is just before it changes its multiplicity. (Color figure online).}
\label{asymf8}
\end{figure*}

\begin{figure*}[!t]
\centering
\resizebox{\hsize}{!}{\includegraphics{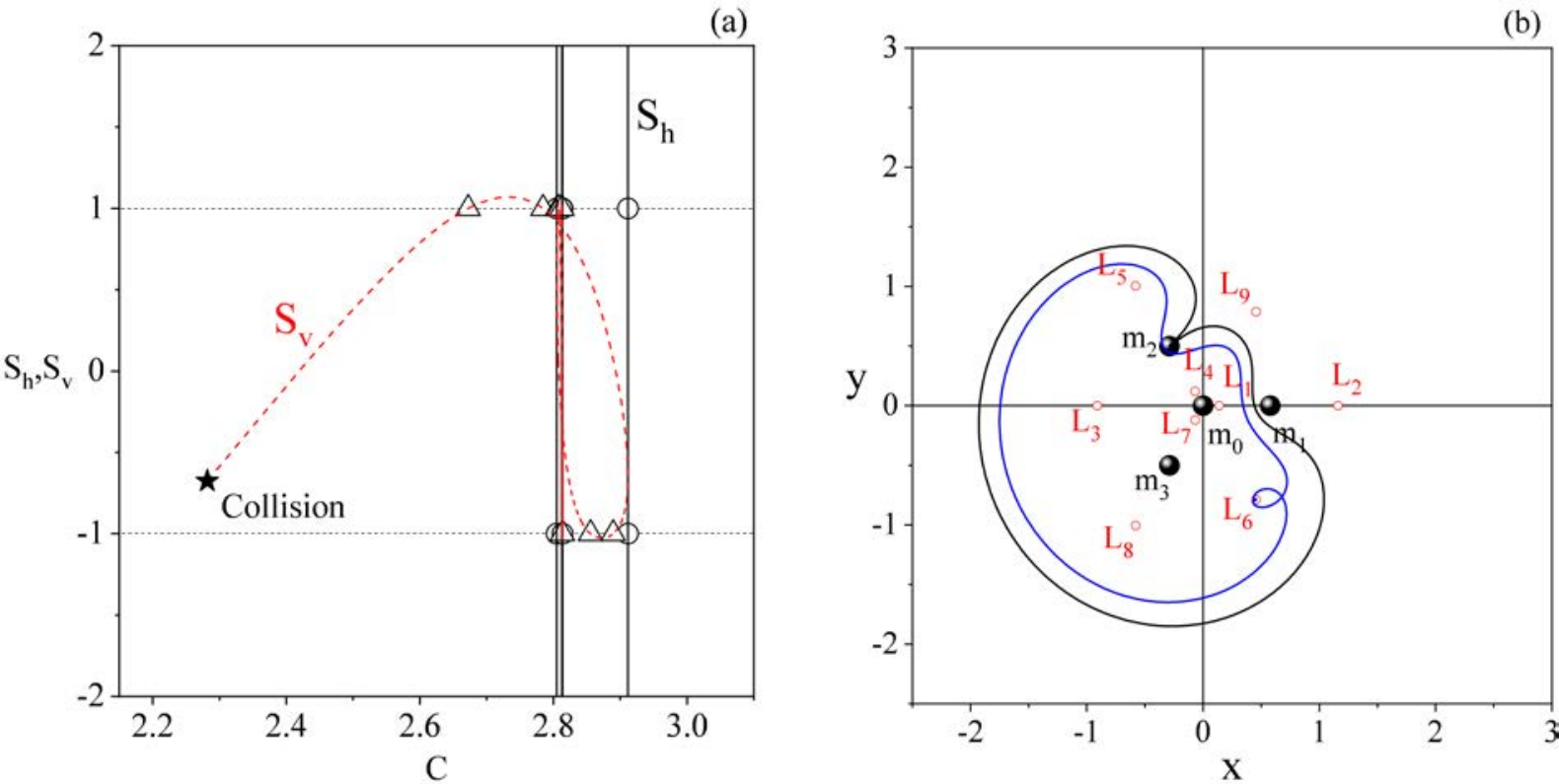}}
\caption{(a-left): The linear stability diagram of the family fam9, for $\beta = 0.05$. (b-right): Two periodic orbits of this family. The black one is just before the fifth body collides with the primary body $P_2$, while the blue orbit is an almost homoclinic asymptotic solution around the equilibrium point $L_6$. (Color figure online).}
\label{asymf9}
\end{figure*}

\begin{figure*}[!t]
\centering
\resizebox{\hsize}{!}{\includegraphics{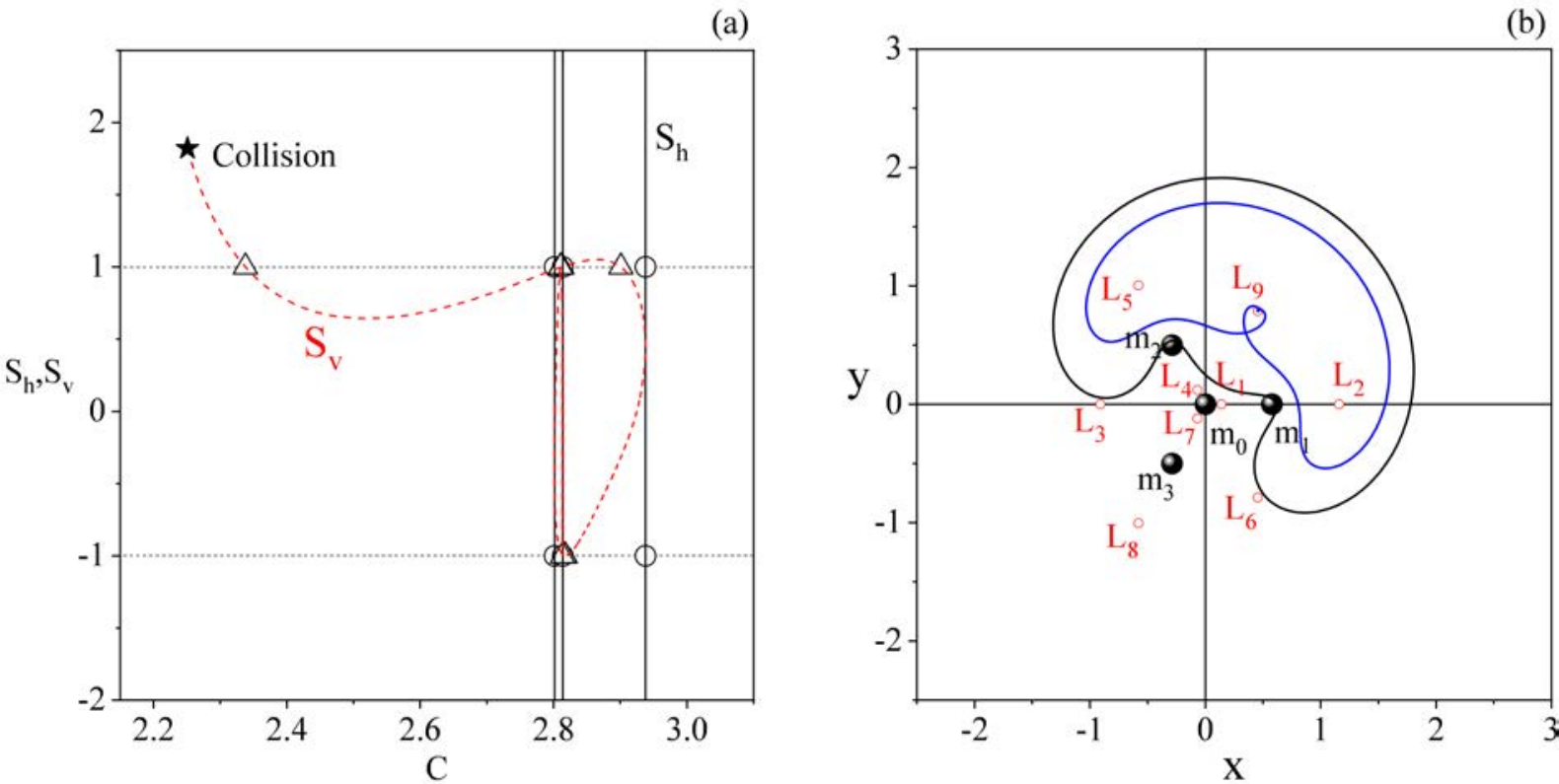}}
\caption{(a-left): The linear stability diagram of the family fam10, for $\beta = 0.05$. (b-right): Two periodic orbits of this family. The black one is just before the test particle collides with the primary bodies $P_1$ and $P_2$, while the blue orbit is an almost homoclinic asymptotic solution around the equilibrium point $L_9$. (Color figure online).}
\label{asymf10}
\end{figure*}

In Fig. \ref{asym2}(a-c) we provide the characteristic curves of the ten families of non-symmetric periodic orbits on the two-dimensional planes $(\dot{x},C)$, $(\dot{y},C)$, and $(T,C)$, for $\beta = 0.05$. In panel (c) it is seen that several of the characteristic curves of the families of non-symmetric periodic orbits tend to the value of the Jacobi constant $C$ which correspond to the equilibrium points $L_3$, $L_6$, and $L_9$, while the time of the period $T$ tends to infinity. This behavior leads us to the conclusion that many of the orbital families have as members non-symmetric asymptotic solutions.

The orbital family fam1 is composed of non-symmetric periodic orbits which circulate around all the primary bodies $P_0$, $P_1$, and $P_3$. In panel (b) of Fig. \ref{asymf1} we show, in green color, a characteristic periodic orbit of the family fam1. With decreasing time the non-symmetric periodic orbits shrink in size and finally they collide with the central primary body $P_0$ (black orbit in the panel (b) of Fig. \ref{asymf1}). On the contrary, with increasing time the size of the orbits grows and finally it leads to collision with the peripheral primary body $P_2$ (blue orbit in the panel (b) of Fig. \ref{asymf1}). All the orbits of the family fam1 are symmetric with respect to the axis of the system which crosses the centers of the primaries $P_0$ and $P_2$, while obviously they are not symmetric, with respect to the horizontal $x$-axis.

The linear stability curves, associated with the stability indices $S_h$ and $S_v$, of the simple non-symmetric periodic obits of family fam1 are depicted in panel (a) of Fig. \ref{asymf1}. It is observed that for the family fam1 stable periodic orbits as well as horizontal and vertical-critical periodic solutions exist. These non-symmetric critical periodic orbits are intersected by periodic orbits of other families, not only on the plane but also in three dimensions.

The orbital family fam2 contains non-symmetric periodic orbits which circulate around the primary bodies $P_0$ and $P_2$. In panel (b) of Fig. \ref{asymf2} we present a characteristic example of periodic solution belonging to family fam2. As the time decreases the non-symmetric periodic orbits of the family fam2 lead to collision with the primary body $P_2$, while with increasing time the periodic solutions lead to collision with the central primary $P_0$. From the shape of the trajectories it becomes evident that these periodic solutions do not display any kind of symmetry, with respect to the symmetry axes of the system.

In panel (a) of Fig. \ref{asymf2} we provide the linear stability diagram of the orbital family fam2, where one can observe that the family contains both stable as well as two horizontal and five vertical-critical periodic orbits. Using a five-pointed star we mark the collision orbits of the family. Specifically, the one with the larger value of the Jacobi constant corresponds to almost collision with the central primary $P_0$, while that with the lowest value of $C$ corresponds to almost collision with the primary body $P_2$.

The non-symmetric periodic orbits of the family fam3 circulate around all primary bodies, with a characteristic loop around the primary $P_3$. With decreasing time the loop around $P_3$ becomes smaller and finally it leads to collision with the same primary. As the time increases, the loop grows in size and tends to the horizontal $x$-axis, which implies that the multiplicity of the orbit is about to change. At the same time the computation of the family is stopped because in this study we are only interested of simple periodic orbits of the system. The orbit in black, shown in panel (b) of Fig. \ref{asymf3}, is an orbit just before the collision with the primary $P_3$, while the orbit, shown in blue, is an orbit adjacent to the $x$-axis, just before the change of the multiplicity of the orbital family. The periodic orbits of the family fam3 are symmetric, with respect to the symmetry axis of the system which passes through the centers of the primary bodies $P_0$ and $P_3$ but they are not symmetric, relative to the horizontal $x$-axis.

The characteristic linear stability curves of the orbital family fam3 are given in panel (a) of Fig. \ref{asymf3}, where it is clearly seen that the family is completely vertical-stable, while a small segment of horizontal-stability is also present. The same orbital family has two horizontal-critical periodic solutions, while vertical-critical periodic orbits do not exist. In the same panel a black five-pointed star indicates the orbit just before the collision with the primary body $P_3$, while using blue five-pointed star we pinpoint the orbit just before the change of the multiplicity of the family.

The orbital family fam4 consists of non-symmetric periodic orbits which circulate around all the primary bodies of the system. This family terminates from both sides with asymptotic solutions. Being more specific, from one side the family leads to homoclinic asymptotic solutions around the equilibrium point $L_9$ (black orbit in panel (b) of Fig. \ref{asymf4}), while from the other side the family leads to heteroclinic asymptotic solutions around the equilibrium points $L_3$ and $L_6$ (blue orbit in panel (b) of Fig. \ref{asymf4}). It should be noted that the asymptotic orbits around the libration points $L_3$ and $L_6$ are obviously not simple periodic orbits, since $L_3$ is a collinear equilibrium point and therefore lies on the horizontal $x$-axis. There the infinite (theoretically) loops around the equilibrium point intersect infinite times the $x$-axis and therefore the multiplicity of the family tends to infinity. Nevertheless, due to the high importance of the asymptotic solutions in a dynamical system, these orbits have been computed and presented in panel (b) of Fig. \ref{asymf4}. The periodic orbits of the family fam4 are symmetric, with respect to the axis of symmetry of the system which passes through the centers of the primaries $P_0$ and $P_3$.

As it is seen from of the linear stability diagram, presented in panel (a) of Fig. \ref{asymf4}, the orbital family fam4 has stable solutions. Furthermore, this family has infinite (theoretically) critical periodic solutions, due to the existence of the asymptotic orbits. In our analysis, we computed more than thirty (30) critical periodic orbits, which are indicated by small circles and triangles in panel (a) of Fig. \ref{asymf4}. The initial conditions of some of these critical periodic orbits are given in Table \ref{tab2}, at the end of this section.

Orbital family fam5 contains simple non-symmetric periodic orbits which circulate around the primary bodies $P_0$ and $P_3$ of the system. From both sides, during its evolution, the orbits of this family lead to collision with the primary $P_2$. A characteristic orbit of the family fam5, just before the collision, is illustrated in panel (b) of Fig. \ref{asymf5}. The solutions of this orbital family are not symmetric, with respect to all of the axis of symmetry of the system.

This orbital family has also stable periodic solutions, according to linear stability diagram given in panel (a) of Fig. \ref{asymf5}. The family fam5 has two horizontal-critical and two vertical-critical periodic orbits, which are indicated by small black circles and triangles, respectively in the linear stability diagram. At the same plot we note, using black five-pointed stars, the periodic orbits, juts before the collision with the primary $P_2$.

The orbital family fam6 is composed of non-symmetric periodic orbits which circulate around of the primary bodies $P_0$ and $P_3$. With decreasing time the orbits shrink and finally they led to collision with the central primary body $P_0$ (orbit in black in panel (b) of Fig. \ref{asymf6}). On the other hand, with increasing time they grow in size and for $t \to \infty$ they lead to asymptotic orbits around the equilibrium points $L_3$ and $L_6$ (orbit in blue in panel (b) of Fig. \ref{asymf6}). The orbits of this family are symmetric, with respect to the axis of symmetry which passes through the centers of the primaries $P_0$ and $P_3$, while they are not symmetric, with respect to the horizontal $x$-axis.

In panel (a) of Fig. \ref{asymf6} we provide the linear stability diagram of the orbital family fam6, where we observe that stable solutions do exist. At the left-hand side of the stability diagram we see the finite number of horizontal and vertical-critical periodic orbits, while at the right-hand side of the same diagram a large number (theoretically there should be infinite number as $t \to \infty$) of both horizontal and vertical-critical periodic orbits is confined, due to the asymptotic solutions of the family.

The members of the orbital family fam7 are non-symmetric periodic orbits which circulate around the equilibrium point $L_2$. This family is one of most interesting orbital families of the system because as the time decreases the family fam7 of the non-symmetric orbits tend to coincide with the family of symmetric periodic orbits which emerge from the libration point $L_2$. The common critical solution (bifurcation) of the symmetric and non-symmetric families is shown, in black color, in panel (b) of Fig. \ref{asymf7}. The initial conditions of the  horizontal-critical periodic orbit are given in Table \ref{tab2}. With increasing time, the orbital family evolves and the corresponding solutions lead to asymptotic orbits, around the equilibrium point $L_9$ (orbit in blue in panel (b) of Fig. \ref{asymf7}). The periodic orbits of the family fam7 are not symmetric with respect to none of the symmetry axes of the system.

The linear stability diagram of the orbital family fam7 is presented in panel (a) of Fig. \ref{asymf7}. In this plot, we indicate the critical orbit which is the bifurcation between the symmetric and the non-symmetric families of periodic orbits. According to the stability diagram, stable periodic orbits are possible for the family fam7. Due to the existence of the asymptotic solutions there exist numerous (theoretically infinite) horizontal and vertical-critical orbits, which are noted using small circles and triangles, respectively.

The orbital family fam8 consists of non-symmetric periodic orbits which circulate around only the central primary body $P_0$. In panel (b) of Fig. \ref{asymf8} the orbit in black is a characteristic orbit of this family. As time decreases, the size of the periodic orbits is reduced and finally they lead to collision with the primary $P_0$. On the contrary, as time increases, the orbits grow in size and finally they intersect again the horizontal $x$-axis, thus changing multiplicity. In the diagram of Fig. \ref{asymf8}b we provide, in blue color, the shape of an orbit, just before it intersects the $x$-axis for more than two times. It is observed that this family contains orbits whose shape is considerably smaller, in relation to the periodic orbits of the other families (note the smaller scale on both axes of the frame in Fig. \ref{asymf8}b). The periodic solutions of the family fam8 are symmetric, with respect to the axis of symmetry which passes through the centers of the primary bodies $P_0$ and $P_2$, but they are not symmetric, with respect to the horizontal $x$-axis.

In panel (a) of Fig. \ref{asymf8} we present the linear stability diagram of the orbital family fam8 from which we deduce that the corresponding periodic orbits are entirely horizontal unstable, while the vast majority of them is vertical stable. The collision orbit with the primary $P_0$ is indicated by a black five-pointed star, while the blue star denotes the periodic orbits, just before the change of the multiplicity. From the same diagram one can see that for the family fam8 we have six vertical-critical periodic solutions (two of them, at the right-hand side of the diagram, are very close to each other), while there is no horizontal-critical periodic orbit.

The orbital family fam9 is composed of non-symmetric periodic obits which circulate around the primary bodies $P_0$ and $P_3$. With decreasing time the family evolves and the corresponding orbits lead to collision with the peripheral primary $P_2$. On the other hand, with increasing time the solutions evolve to asymptotic periodic orbits around the libation point $L_6$. In panel (b) of Fig. \ref{asymf9} we depict, in black color, a characteristic orbit of this family, just before the collision with the primary $P_2$, while in the same plot, with blue color, we provide an almost asymptotic orbit to $L_6$. All the periodic orbits of the family fam9 are non-symmetric, with respect to all of the axes of symmetry of the system.

\begin{figure*}[!t]
\centering
\resizebox{\hsize}{!}{\includegraphics{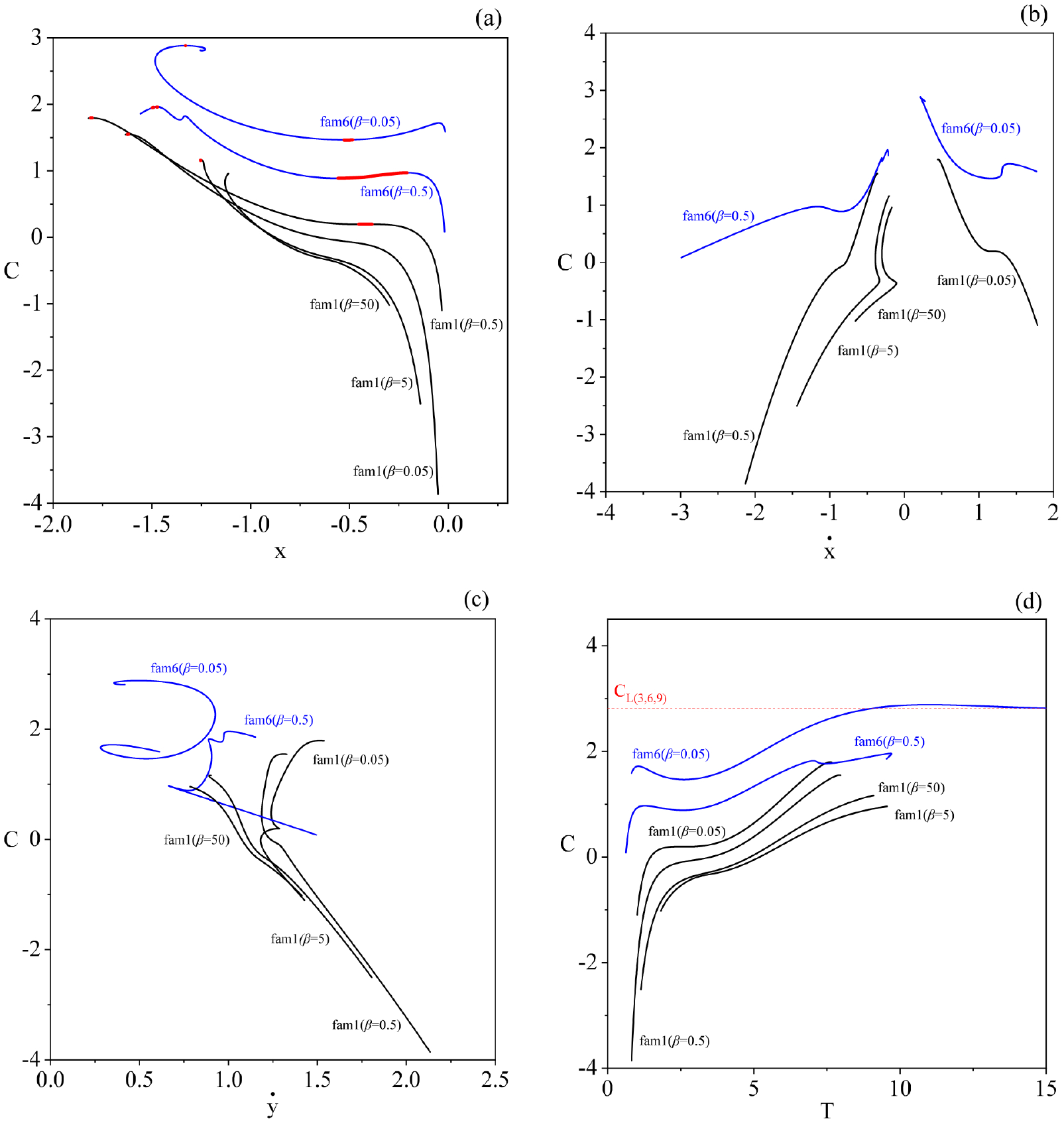}}
\caption{The black solid lines are the characteristic curves on the $(x,C)$, $(\dot{x},C)$, $(\dot{y},C)$ and $(T,C)$ of the family fam1 of the non-symmetric periodic orbits, for $\beta = 0.05, 0.5, 5$ and 50, respectively. Same presentation of the family fam6 (blue lines), for $\beta = 0.05$ and 0.5. In the $(x,C)$ plane we indicate the stable non-symmetric periodic solutions using red arcs. (Color figure online).}
\label{charf}
\end{figure*}

\begin{figure*}[!t]
\centering
\resizebox{\hsize}{!}{\includegraphics{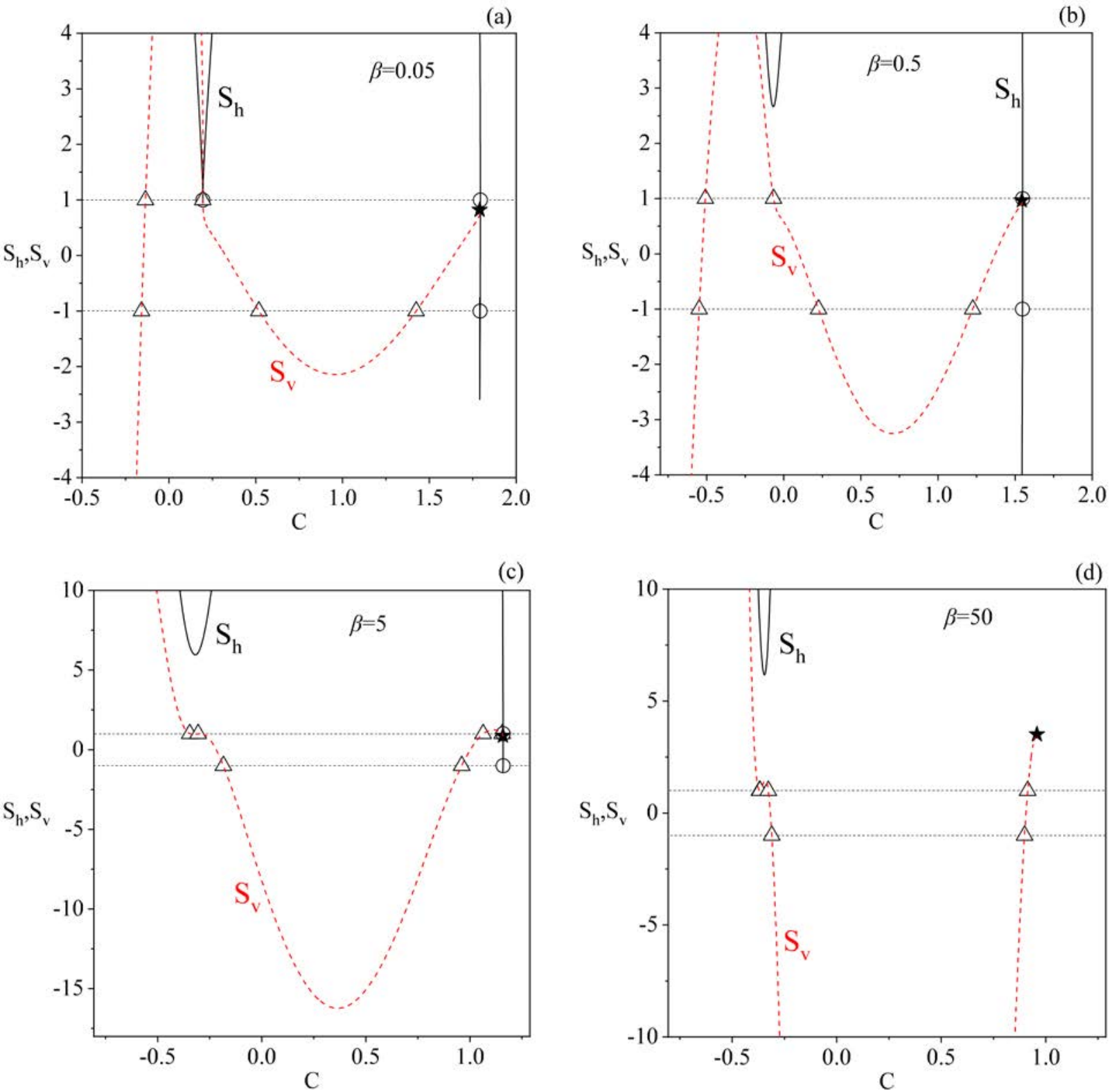}}
\caption{Stability diagrams of the same family fam1 of the non-symmetric periodic orbits, for $\beta = 0.05, 0.5, 5$ and 50, respectively. The small black five-pointed stars denote collision orbits. (Color figure online).}
\label{sff}
\end{figure*}

\begin{figure*}[!t]
\centering
\resizebox{\hsize}{!}{\includegraphics{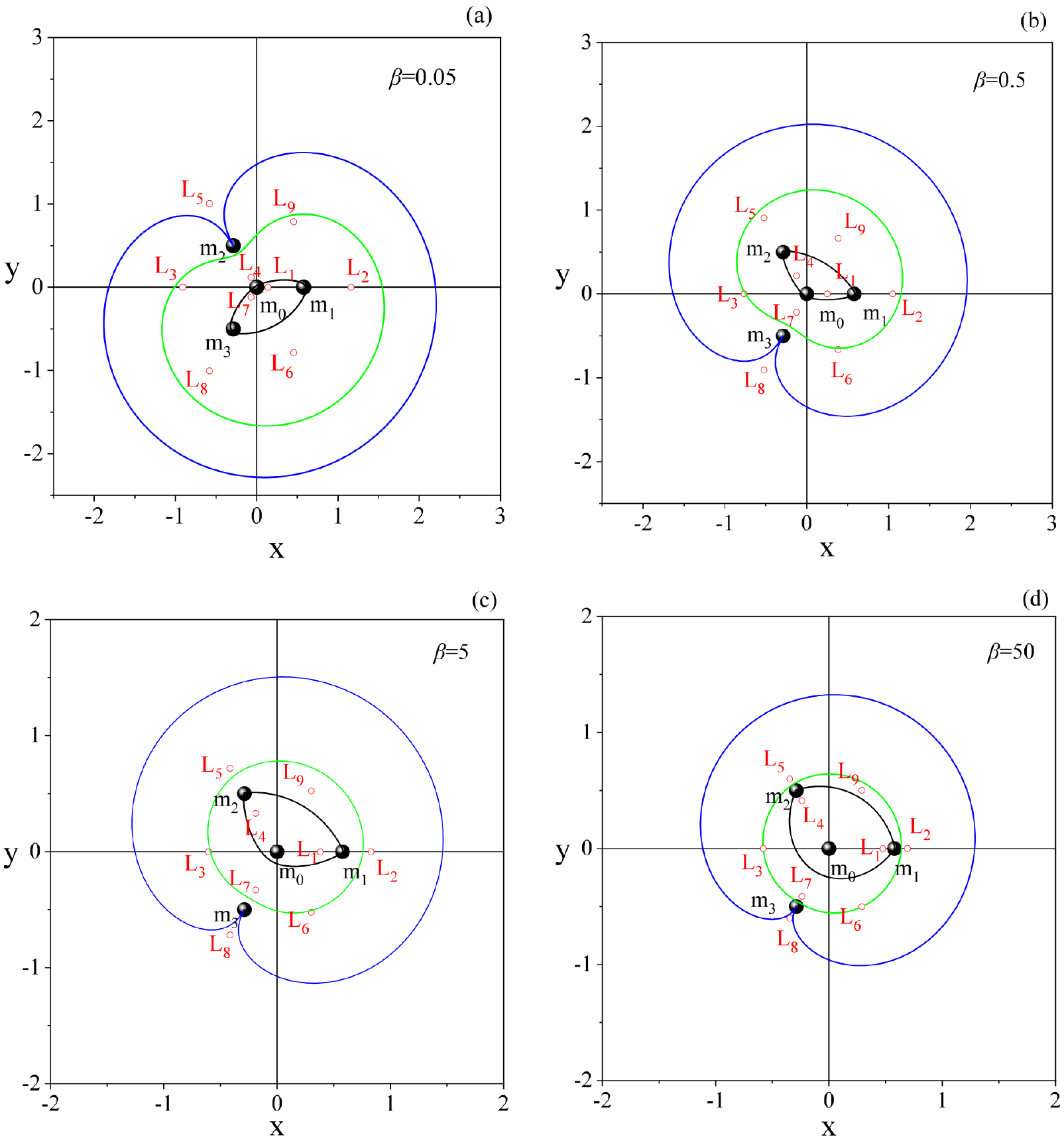}}
\caption{Non-symmetric periodic orbits of the same family fam1 but for different value of the mass parameter $\beta$, (a-upper left): $\beta = 0.05$; (b-upper right): $\beta = 0.5$; (c-lower left): $\beta = 5$; (d-lower right): $\beta = 50$. (Color figure online).}
\label{orbsf}
\end{figure*}

The linear stability diagram, corresponding to orbital family fam9, is presented in panel (a) of Fig. \ref{asymf9}. One can observe that the family contains stable solutions with (theoretically) infinite critical periodic orbits, due to the loops of the orbits which tend asymptotically to the equilibrium point $L_6$. A periodic orbit, just before the collision with the primary $P_2$, is indicated using a five-pointed star in the diagram of Fig. \ref{asymf9}a.

Finally, the orbital family fam10 consists of non-symmetric periodic orbits which circulate outside the primaries of the system, while they pass close to the primary bodies $P_1$ and $P_2$. From one side (when $t \to 0$) the family leads to collision with the bodies $P_1$ and $P_2$ (orbit in black in panel (b) of Fig. \ref{asymf10}), while from the other side (when $t \to \infty$) the family tends to an asymptotic solution around the equilibrium point $L_9$ (orbit in blue in panel (b) of Fig. \ref{asymf10}). All the periodic orbits of this family are symmetric, with respect to the axis of symmetry which passes through the centers of the primaries $P_0$ and $P_3$, while they are not symmetric, with respect to the horizontal $x$-axis.

\begin{table*}[!t]
\caption[]{Initial conditions, value of the Jacobi constant, period and linear stability of some horizontal or vertical-critical non-symmetric simple periodic orbits of the ten orbital families, for $\beta = 0.05$}
\centering
\begin{tabular}{lrrrrl}
\hline\noalign{\smallskip}
\multicolumn{1}{c}{Family} & \multicolumn{1}{c}{$x_{0}$} & \multicolumn{1}{c}{$\dot{x}_0$}
& \multicolumn{1}{c}{$C$} & \multicolumn{1}{c}{$T$} & \multicolumn{1}{c}{Stability} \\  \noalign{\smallskip}\hline\noalign{\smallskip}
fam1  & $-0.4530915914$    &  $1.1611825251$ &    0.1950438943 &  2.7000473275 & $S_h=+1$ \\
      & $-0.4787482214$    &  $1.1465180561$ &    0.1953848942 &  2.8046790004 & $S_v=+1$ \\
fam2  & $-0.3425181009$    & $-0.0809920294$ &    3.2634111433 &  7.7165292968 & $S_h=-1$ \\
      & $-0.3464774446$    & $-0.1298370038$ &    3.2335455547 &  6.8871511570 & $S_v=+1$ \\
fam3  & $-2.1686273165$    & $-0.3575394816$ &    0.9071377868 &  9.1907941260 & $S_h=-1$ \\
fam4  & $-2.2228626797$    &  $0.1125792676$ &    2.8185187262 & 15.6021185990 & $S_h=+1$ \\
      & $-1.1821375517$    & $-0.1740532194$ &    2.9105389711 & 14.6827720539 & $S_v=-1$ \\
fam5  & $-1.8965269060$    &  $0.1552291395$ &    1.9885987682 &  7.7500736031 & $S_h=+1$ \\
      & $-1.8664828264$    &  $0.1775189089$ &    2.0304809490 &  7.9299792287 & $S_v=-1$ \\
fam6  & $-0.4839942689$    &  $1.1976251816$ &    1.4659656224 &  2.4809193898 & $S_h=-1$ \\
      & $-1.4800292654$    &  $0.2743715248$ &    2.7133466436 &  8.1665105354 & $S_v=-1$ \\
fam7  &  $0.6964066284$    &  $0.0000000000$ &    2.0760662587 &  5.6402575055 & $S_h=+1$ \\
      &  $0.7989687584$    & $-0.2692125707$ &    2.8297659402 &  8.6452549930 & $S_h=-1$ \\
      &  $0.7952918939$    & $-0.2618104829$ &    2.8168688164 & 10.2403969301 & $S_v=+1$ \\
fam8  &  $0.0725433078$    &  $0.0799256157$ &    3.4695481261 &  1.7192225562 & $S_v=-1$ \\
fam9  &  $0.3449692834$    &  $0.2535837118$ &    2.8048471780 & 13.6675829779 & $S_h=+1$ \\
      &  $0.3973562867$    &  $0.5104211783$ &    2.6724387012 &  8.7667962789 & $S_v=+1$ \\
fam10 &  $0.8460308024$    & $-0.2991855350$ &    2.9377204399 &  9.9078963945 & $S_h=-1$ \\
      &  $0.8089040894$    & $-0.2675791724$ &    2.8176473226 & 11.7367651088 & $S_v=-1$ \\
\noalign{\smallskip}\hline
\end{tabular}
\label{tab2}
\end{table*}

Looking at the linear stability diagram, given in panel (a) of Fig. \ref{asymf10}, it becomes evident that stable periodic solutions exist for the orbital family fam10. In the same diagram, we mark, several of the critical periodic orbits (theoretically there should be infinite such orbits), using circles for the horizontal-critical ones and triangles for the vertical-critical one, while with a five-pointed star we indicate a periodic orbit, just before collision with the primaries $P_1$ and $P_2$.

In Table \ref{tab2} we provide, for each of the ten orbital families, the initial conditions, the value of the Jacobi constant $C$, the period $T$, as well as the stability of one horizontal and one vertical-critical periodic orbit. For all these non-symmetric periodic solutions we have that $\dot{y_0} > 0$. For the orbital family fam3 we do not give any vertical-critical periodic orbit, because fam3 is entirely vertical stable (see panel (a) of Fig. \ref{asymf3}) and therefore critical orbits with $S_v = \pm 1$ simply do not exist. Similarly, we dot not provide any horizontal-critical periodic orbit for the family fam8, because this orbital family is entirely horizontal stable (see panel (a) of Fig. \ref{asymf8}). The orbital family fam3 is intersected with a family of symmetric periodic orbits, which emerge in the vicinity of the collinear equilibrium point $L_2$ of the system. In Table \ref{tab2}, apart from one horizontal and one vertical-critical solution of family fam3 we also give the common periodic orbits which belongs to both families (being a symmetric solution it has $\dot{x_0} = 0$). Moreover, it should also be noted that in Table \ref{tab2}, in contrast with Table \ref{tab1}, the provided infirmation has increased accuracy of ten significant decimal figures. This is because for the symmetric solutions we could stop the computations after time equal to half of the period $(T/2)$ of the orbits, while on the other hand for the non-symmetric solutions the computations require a time interval equal to the full period $T$ of the orbits.

\subsubsection{Non-symmetric periodic solutions for $\beta = \{0.5, 5, 50 \}$}
\label{nsp2}

In order to investigate how the mass parameter $\beta$ influences the properties of the families of the non-symmetric periodic orbits we study the evolution of the orbital families fam1 and fam6, when $\beta = \{0.5, 5, 50 \}$.

The characteristic curves of the orbital families fam1 and fam6 on the $(x,C)$ plane, for $\beta = \{0.05, 0.5, 5, 50 \}$, are presented in panel (a) of Fig. \ref{charf}. The influence of the mass parameter on the evolution of the families becomes evident, while it should be pointed out that the family fam6 does not exist when $\beta = 5$ and $\beta = 50$. In the same diagram we also provide, with red color, the horizontal linear stability of the periodic orbits. Therefore, the red arcs in every characteristic curve denote horizontal stable non-symmetric periodic orbits of the families fam1 and fam6. All the orbital families have stable periodic solutions apart from the family fam1 for $\beta = 50$.

As it was explained in the previous subsection, the presentation of the non-symmetric periodic solutions on the $(x,C)$ plane is not sufficient, since it does not provide any information regarding the initial conditions of the velocities of the test particle (fifth body). For this reason, in panels (b) and (c) of Fig. \ref{charf} we illustrate the characteristic curves of the orbital families fam1 and fam6 on the $(\dot{x},C)$ and $(\dot{y},C)$ planes, respectively. Similarly, in panel (d) of the same figure we see the parametric evolution of the Jacobi constant $C$, as a function of the period $T$ of the orbits, on the $(T,C)$ plane.

For determining the influence of the mass parameter on the linear stability of the periodic orbits we present in Fig. \ref{sff}(a-d) the corresponding stability diagrams of the family fam1, for the four values of the mass parameter. This family has horizontal and vertical stable periodic solutions, for all values of $\beta$, apart from the case with $\beta = 50$, where the family fam1 is entirely horizontal unstable, with small segments of vertical stability. With increasing value of the mass parameter the intervals with horizontal linear stability decrease. For all studied values of $\beta$ the family fam1 has horizontal and vertical-critical periodic solutions, apart from the case with $\beta = 50$, where only vertical-critical periodic solutions exist. Nevertheless, the vertical stability of the orbits is also influenced by the mass parameter. This is true because as the value of $\beta$ increases, the intervals of the vertical stability decrease.

In Fig. \ref{orbsf}(a-b) we show characteristic non-symmetric periodic orbits of the orbital family fam1, for $\beta = \{0.05, 0.5, 5, 50 \}$. For all the values of the mass parameter, the family fam1 is composed by retrograded non-symmetric periodic orbits, which circulate around the primary bodies $P_0$, $P_1$, and $P_3$. With decreasing time, the orbits of this family, for all values of $\beta$, are reduced and they approach the three primaries, while they finally they lead to collision with the primary $P_0$. On the contrary, with increasing time, the periodic orbits grow in size and finally they lead to collision with the primary $P_2$ (for $\beta = 0.05$) and with the primary $P_3$ (for $\beta = \{0.5, 5, 50 \}$). This difference, regarding to which primary they collide with increasing time, is due to the initial value of the velocity $\dot{x}$. In particular, for $\beta = 0.05$ the initial value of $\dot{x}$ is positive, while for larger values of the mass parameter it becomes negative $(\dot{x_0} < 0)$ (see panel (b) of Fig. \ref{charf}, where the signs of $\dot{x}$ are clearly visible). This change on the sign of the initial velocity is responsible for the differences on the shape of the periodic orbits, for the different values of the mass parameter.

\section{Concluding remarks}
\label{conc}

In this work the orbital dynamics of the planar circular restricted five-body problem have been numerically investigated. Through the numerical integration of large sets of initial conditions we managed to classify them into three main categories: (i) bounded (regular or chaotic) (ii) escaping and (iii) collision orbits. The influence of the mass parameter $\beta$ on the orbital dynamics and on the degree of fractality of the phase space has been determined. The networks of the families of both symmetric and non-symmetric periodic orbits have been explored, by using the grid method. Moreover, the parametric evolution of the horizontal and vertical linear stability (at first approximation) of the periodic solutions, as a function of the mass parameter, has also been studied.

A large number of families of both symmetric and non-symmetric simple periodic orbits of the system have been found, while an analytic description of the most important outcomes, for four values of the mass parameter, have been presented. We located periodic orbits which circulate around one, two, three or even four primary bodies or equilibrium points. In addition, homoclinic and heteroclinic asymptotic solutions, around the libration points, have been found. These solutions have useful practical applications, since they circulate around an equilibrium point for infinite time $(t \to \infty)$. At this point, it should be noted, that the majority of these solutions are in fact stable periodic solutions. By comparing several results of the periodic solutions, for different values of the mass parameter, we concluded that the parameter $\beta$ strongly influences the periodic solutions of the system.

As far as we know, there are no previous related studies containing such systematic and thorough numerical analysis on the orbital dynamics of the planar restricted five-body problem. More precisely, this is the first time where results regarding the nature as well as the linear stability of non-symmetric periodic solutions on the restricted five-body problem are presented. On this basis, all the contained outcomes are novel and add to our existing knowledge on the orbital properties of the dynamical systems in the field of the few-body problem.

For classifying the initial conditions of the orbits on the two-dimensional planes we used a double precision Bulirsch-Stoer \verb!FORTRAN 77! code \cite{PTVF92}. The required CPU time, per grid of initial conditions, was varying between half and 3 days, using an Intel$^{\circledR}$ Quad-Core\textsuperscript{TM} i7 3.9 GHz PC. The latest versions 11.3 and 2018b of the softwares Mathematica$^{\circledR}$ \cite{W03} and Origin, respectively were used for developing all the graphical illustration of the paper.

In this article, we focused on the orbital dynamics of the planar version (two degrees of freedom) of the restricted five-body problem. An undeniably challenging task, for a future work, would be to unveil the orbital dynamics of the full system of three degree of freedom.

\section*{Acknowledgments}
\footnotesize

Our warmest thanks go to two anonymous referees for the careful reading of the manuscript as well as for all the apt suggestions and comments which allowed us to improve both the quality and the clarity of the paper.

\end{document}